\documentclass[aps,prl,superscriptaddress,twocolumn,showpacs]{revtex4-1}

\usepackage[utf8]{inputenc}
\usepackage[T1]{fontenc}
\usepackage[german,english]{babel}
\usepackage{amsmath}
\usepackage{mathtools}
\usepackage{amssymb}
\usepackage{amsthm}
\usepackage{tensor}
\usepackage{microtype} %microtipografia ottimale
\usepackage{quoting} %citazione
\quotingsetup{font=small}
\usepackage{tensor}
\usepackage{graphicx}
\usepackage{fancyhdr}
\usepackage{braket}
\usepackage{bm}
\usepackage{comment}
\usepackage{booktabs}
\usepackage{eurosym} % € = \euro
\usepackage{braket}
\usepackage{bbold} % identity operator
\usepackage{emptypage}
\usepackage{epstopdf}
\usepackage{verbatim}
\usepackage{enumitem}
\usepackage{float}
\usepackage{chemformula}

\newcommand{\beq}{\begin{equation}}
\newcommand{\eeq}{\end{equation}}

\renewcommand{\sim}{\thicksim}

\usepackage[bookmarks=true,colorlinks,linkcolor=red,urlcolor=blue,citecolor=blue]{hyperref} % per ultimo!
\begin{document}

\title{Reaction-limited quantum reaction-diffusion dynamics}
\author{Gabriele Perfetto}
\thanks{gabriele.perfetto@uni-tuebingen.de}
\affiliation{Institut f\"ur Theoretische Physik, Eberhard Karls Universit\"at T\"ubingen, Auf der
Morgenstelle 14, 72076 T\"ubingen, Germany.}
\author{Federico Carollo}
\affiliation{Institut f\"ur Theoretische Physik, Eberhard Karls Universit\"at T\"ubingen, Auf der
Morgenstelle 14, 72076 T\"ubingen, Germany.}
\author{Juan P. Garrahan}
\affiliation{School of Physics, Astronomy, University of Nottingham, Nottingham, NG7 2RD, UK.}
\affiliation{Centre for the Mathematics, Theoretical Physics of Quantum Non-Equilibrium Systems,
University of Nottingham, Nottingham, NG7 2RD, UK}
\author{Igor Lesanovsky}
\affiliation{Institut f\"ur Theoretische Physik, Eberhard Karls Universit\"at T\"ubingen, Auf der
Morgenstelle 14, 72076 T\"ubingen, Germany.}
\affiliation{School of Physics, Astronomy, University of Nottingham, Nottingham, NG7 2RD, UK.}
\affiliation{Centre for the Mathematics, Theoretical Physics of Quantum Non-Equilibrium Systems,
University of Nottingham, Nottingham, NG7 2RD, UK}
%\date{\today} %per non stampare la data, decommentare e parentesi vuote

\begin{abstract}
We consider the quantum nonequilibrium dynamics of systems where fermionic particles coherently hop on a one-dimensional lattice and are subject to dissipative processes analogous to those of classical reaction-diffusion models. Particles can either annihilate in pairs, $A+A \to \emptyset$, or coagulate upon contact, $A+A \to A$, and possibly also branch, $A \to A+A$. In classical settings, the interplay between these processes and particle diffusion leads to critical dynamics as well as to absorbing-state phase transitions. Here, we analyze the impact of coherent hopping and of quantum superposition, focusing on the so-called reaction-limited regime. Here, spatial density fluctuations are quickly smoothed out due to fast hopping, which for classical systems is described by a mean-field approach. By exploiting the time-dependent generalized Gibbs ensemble method, we demonstrate that quantum coherence and destructive interference play a crucial role in these systems and are responsible for the emergence of locally protected dark states and collective behavior beyond mean-field. This can manifest both at stationarity and during the relaxation dynamics. Our analytical results highlight fundamental differences between classical nonequilibrium dynamics and their quantum counterpart and show that quantum effects indeed change collective universal behavior.
\end{abstract}

\maketitle

\textbf{Introduction.}---
In reaction-diffusion (RD) models classical reactants, or particles, are transported by diffusion and react when they meet, see, e.g., Refs.~\cite{vladimir1997nonequilibrium,hinrichsen2000non,henkel2008non}. These are paradigmatic non-equilibrium systems displaying universal dynamical properties and stationary-state transitions from fluctuating phases to absorbing states, i.e., states that once reached cannot be left. In one dimension, in particular, spatial fluctuations of the particle number dominate the kinetics and both exact analytical results \cite{toussaint1983particle,Spouge1988,Privman1994,torney1983diffusion,Racz1985,Tasaki1988} and dynamical field-theory renormalization calculations \cite{doi1976stochastic,tauber2005applications,QFT_RD_1998,peliti1986renormalisation,peliti1985path,tauber2014critical,grassberger1980fock} have shown that the dynamical critical behavior is universal and it is not captured by the mean-field approximation. This is especially true in the \textit{diffusion-limited} regime, i.e., when the diffusive mixing of the particles is not too strong \cite{ovchinnikov1978role,fluctuationseffects,toussaint1983particle,redner1984scaling,Privman1994}. In the opposite \textit{reaction-limited} regime, where the diffusive motion is fast, the density of reactants rapidly uniformize (leading to the alternative name of well-stirred-mixture approximation) and one recovers mean-field results \cite{hinrichsen2000non,vladimir1997nonequilibrium,tauber2005applications,Redner1984,fastdiffusion1992}.

Quantum effects can alter the universal properties of absorbing-state phase transitions. This has been shown for Markovian open quantum systems \cite{griessner2006,diehl2008,kraus2008,diehl2011,tomadin2011,bardyn2013,perez-espigares2017,buca2020,carollo2022}, for systems with kinetic constraints \cite{lesanovsky2013,olmos2014,everest2016,marcuzzi2016,buchhold2017,gutierrez2017,roscher2018,carollo2019,gillman2019,gillman2020,wintermantel2020,helmrich2020,nigmatullin2021,kazemi2021,carollo2022quantum} and for the quantum contact process \cite{carollo2019,jo2021}. 
Quantum dissipative RD spin chains, where the diffusive motion is replaced by coherent hopping, have been investigated in Ref.~\cite{RDHorssen}. However, results in this and other works are limited to small systems, due to the complexity of the numerical simulation of many-body quantum dynamics. As a consequence, very little is known about the impact of quantum effects on universal aspects of RD dynamics and on absorbing-state phase transitions.

\begin{figure}[t]
    \centering
    \includegraphics[width=1\columnwidth]{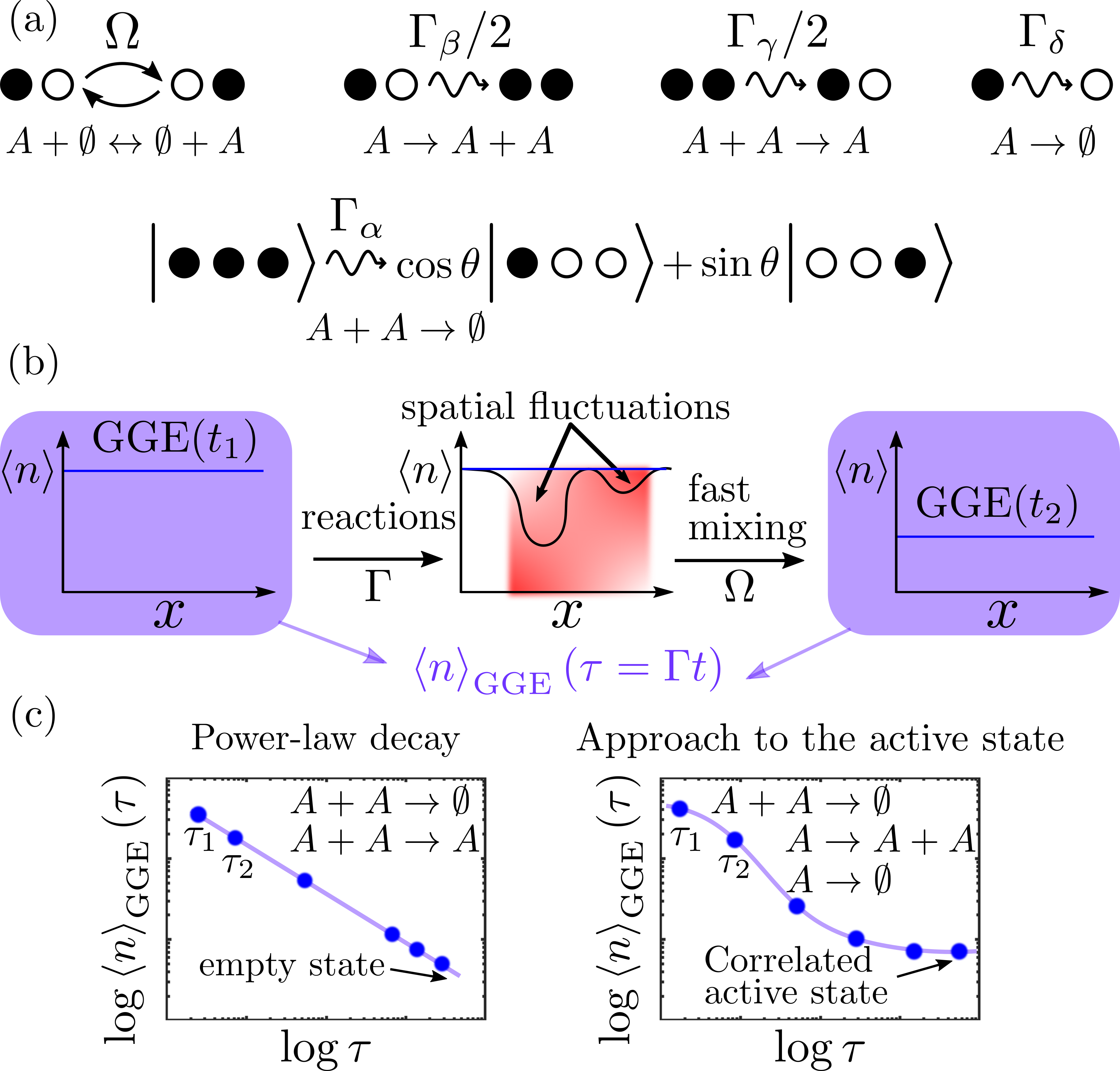}
    \caption{\textbf{Quantum RD dynamics in the reaction limited regime.} (a) Quantum chain with sites that can either be occupied by a fermion, $\ket{\cdots \bullet_j \cdots}$, or empty $\ket{\cdots \circ_j \cdots}$. Particles can hop between nearest-neighboring sites with hopping rate $\Omega$, Eq.~\eqref{eq:free_fermion_Hamiltonian}. Dissipation consists of irreversible reactions at rate $\Gamma_{\nu}$, Eqs.~\eqref{eq:annihilation}-\eqref{eq:death}.
    The parameter $\theta$ controls coherent superposition from pair annihilation events. (b) In the reaction-limited regime, $\Gamma \ll \Omega$, reaction dynamics is slow and takes place on the timescale $\sim\Gamma^{-1}$. Fast hopping rapidly smooths out spatial fluctuations (highlighted in red), due to local reactions, and the state of the systems is described by a homogeneous GGE($\tau$) (blue horizontal lines) at any rescaled time $\tau=\Gamma t$. (c) The total particle density $\braket{n}_{\mathrm{GGE}}(\tau)$ decays algebraically in rescaled time $\tau$ (blue points) for annihilation or coagulation with exponent dependent on initial state coherence. When branching is included an absorbing-state phase transition to an active, finite density of particles, state can occur. The latter displays correlation when $\theta \neq 0,\pi/2$.} 
    \label{fig:sketch_RD_GGE}
\end{figure}

In this manuscript, we make progress in this direction, deriving exact analytical results for the case of reaction-limited open quantum RD processes in fermionic chains.
We consider a series of prototypical reaction processes, such as annihilation $A+A\to \emptyset$, coagulation $A+A\to A$, and branching $A \to A+A$ (see Fig.~\ref{fig:sketch_RD_GGE}), and show that the reaction-limited regime of quantum RD models cannot be described within a mean-field approach, in stark contrast to the classical settings. 
We demonstrate that the presence of quantum effects strongly affects the approach to stationarity and the stationary state itself. 
For annihilation and coagulation, the density of particles features an algebraic (power-law) decay. This power law changes and may deviate from the mean-field predictions when the initial state of the dynamics features quantum coherence. %In general this leads to a slowing down of convergence towards the empty state. 
In the presence of the branching process, quantum RD models display an absorbing-state phase transition. 
Here, annihilation processes that couple to coherent superpositions of adjacent particle pairs  lead to the emergence of dark states which are locally protected against dissipation. These local dark states, which are not captured by the mean-field approach, establish quantum correlations between fermionic particles. 

Our analysis is performed by exploiting the time-dependent generalized Gibbs ensemble method (TGGE) \cite{tGGE1,tGGE2,tGGE3,tGGE4}, which naturally leads to large-scale Boltzmann-like equations. The latter provides an exact description for the reaction-limited regime in the thermodynamic limit. Our analytical findings show that quantum effects lead to rich non-equilibrium behavior, significantly different from that of classical systems. Our results connect to the physics of cold atoms, where losses are of central experimental \cite{lossexp0,lossexp1,lossexp2,lossexp3,lossexp4,lossexp5,lossexp6,lossexp7} and theoretical \cite{lossth7,lossth8,lossth1,lossth2,lossth3,lossth4,lossth5,lossth6} relevance.

\textbf{Quantum reaction-diffusion models}---
We consider fermionic quantum chains of length $L$. Each site $j$ can be either occupied %by a fermion 
$n_j \ket{\cdots \bullet_{j} \cdots}=\ket{\cdots \bullet_j \cdots}$ or empty $n_j \ket{\cdots \circ_j \cdots}=0$, where $n_j=c_j^{\dagger}c_j$ and the operators $c_j,c_j^{\dagger}$ obey the fermionic anticommutation relations $\{c_j,c_{j'}^{\dagger}\}=\delta_{j,j'}$. The fermionic statistics prevents double occupancy of lattice sites, typically assumed in RD classical models \cite{hinrichsen2000non,henkel2008non,vladimir1997nonequilibrium}. The dynamics is ruled by the quantum master equation \cite{gorini1976,lindblad1976,breuer2002} ($\hbar=1$ henceforth) 
\begin{equation}
\dot{\rho}(t)=-i[H,\rho(t)]+\mathcal{D}[\rho(t)]. 
\label{eq:master_equation}   
\end{equation}
Here, we assume that the diffusive motion of the particles in classical RD models is replaced by coherent hopping, which is accounted for by the quantum Hamiltonian
\begin{equation}
H=-\Omega\sum_{j=1}^{L}(c_j^{\dagger} c_{j+1}+c_{j+1}^{\dagger}c_j) \, ,
\label{eq:free_fermion_Hamiltonian}  
\end{equation}
with $\Omega$ the hopping rate [cf.~Fig.~\ref{fig:sketch_RD_GGE}(a)]. Such Hamiltonian is diagonalized 
with Fourier-space fermionic operators $\hat{c}_k,\hat{c}_k^{\dagger}$, where $k$ is the quasi-momentum, and the number operators $\hat{n}_k=\hat{c}_k^{\dagger}\hat{c}_k$ \cite{franchini2017introduction}. It conserves the total number $N=\sum_j n_j=\sum_{k}\hat{n}_k$ of particles: $[H,N]=0$. 
The irreversible reaction processes are encoded in the dissipator $\mathcal{D}$. It takes the (Lindblad) form \cite{gorini1976,lindblad1976,breuer2002}
\begin{equation}
\mathcal{D}[\rho]=\sum_{j,\nu} \left[L_{j}^{\nu}\rho {L_{j}^{\nu}}^\dagger-\frac{1}{2}\left\{{L_{j}^{\nu}}^\dagger L_{j}^{\nu},\rho \right\}\right],
\label{eq:dissipator}    
\end{equation}
where $L_j^{\nu}$ are local jump operators. We consider four different reactions, labelled by the parameter $\nu$, which are sketched in Fig.~\ref{fig:sketch_RD_GGE}(a). The first is binary annihilation, $A +A \to \emptyset$, of a pair of neighboring particles (rate $\Gamma_{\alpha}$), which is described by the jump operators
\begin{equation}
L_j^{\alpha}=L_j^{\alpha}(\theta)=\sqrt{\Gamma_{\alpha}}c_j(\cos\theta \, c_{j+1}-\sin\theta \, c_{j-1}).
\label{eq:annihilation}
\end{equation}
The sum of the two terms, whose balance is controlled by the angle $\theta \in [0,\pi)$, allows for the possibility that interference between two quantum mechanical amplitudes contributes to the pair annihilation process. Such structure naturally emerges in the Bose-Hubbard model subject to strong two-body losses. In this limit, the model can be mapped to free fermions \eqref{eq:free_fermion_Hamiltonian} with weak, $\Gamma_{\alpha}\ll \Omega$, two-body losses \eqref{eq:annihilation}, as shown in Refs.~\cite{lossexp0,lossth7,lossth1}. The classical-incoherent annihilation process is recovered for $\theta=0,\pi/2$. The second reaction is coagulation, $A+A \to A$, of a particle upon meeting a neighbouring one (rate $\Gamma_{\gamma}/2$), with jump operators 
\begin{equation}
L_{j}^{\gamma\pm}=\sqrt{\Gamma_{\gamma}/2} \, c_j n_{j\pm 1}. \label{eq:coagulation}
\end{equation}
The third reaction is one-body annihilation, $A \to \emptyset$, (rate $\Gamma_\delta$) with jump operators
\begin{equation}
L_j^{\delta} = \sqrt{\Gamma_{\delta}}\,c_j\, .
\label{eq:death}
\end{equation}    

These three reactions break number conservation and, due to continued particle loss, drive the system towards an absorbing state devoid of particles. To establish a non-trivial steady state, 
we consider a fourth reaction, namely branching, $A \to A+A$. This process allows for creation of a particle in the neighborhood of an occupied site (rate $\Gamma_{\beta}/2$) 
\begin{equation}
L_j^{\beta\pm}=\sqrt{\Gamma_{\beta}/2}\, c_j^{\dagger}n_{j\pm 1} .
\label{eq:branching}
\end{equation}  
The competition between the branching process and one-body annihilation (as in the contact process \cite{hinrichsen2000non,henkel2008non}) gives rise to a nonequilibrium absorbing-state phase transition, from the empty state to a stationary active one with finite density of particles. Coagulation \eqref{eq:coagulation} and branching \eqref{eq:branching} can be experimentally implemented in the facilitation regime \cite{Garrahan2014facilitation} of cold-atomic gases dressed with Rydberg interactions \cite{EXPKConstraint2016,EXPabs2017,EXP2021epidemic}. For convenience, in the following when multiple reactions are present, we rescale rates as $\Gamma_{\nu}=\Gamma \nu$, so that $\Gamma$ sets the timescale of the  dissipation, while the dimensionless parameters $\alpha$, $\beta$, $\gamma$ and $\delta$ encode the relative strength of the reactions [see Fig.~\ref{fig:sketch_RD_GGE}(b)-(c)]. 

There are two important timescales in the dynamics: the reaction time $\sim \Gamma^{-1}$, which gives the typical time needed for neighbouring particles to react, and the hopping time (or diffusion time in classical RD models) $\sim \Omega^{-1}$, which sets the timescale for two reacting particles to meet. In classical settings \cite{hinrichsen2000non,vladimir1997nonequilibrium}, the dynamics qualitatively changes depending on the ratio $\Gamma/\Omega$. The regime with $\Gamma/\Omega \gg 1$ is named \textit{diffusion limited} as the propagation of particles is the limiting factor for reactions to occur. In this regime, spatial fluctuations are relevant and in one dimension the total particle density $\braket{n}(t)=\braket{N}(t)/L$ decays algebraically as 
$\braket{n}(t) \sim (\Omega t)^{-1/2}$ \cite{toussaint1983particle,Spouge1988,Privman1994,torney1983diffusion,Racz1985,Tasaki1988,ovchinnikov1978role,fluctuationseffects,toussaint1983particle,redner1984scaling,Privman1994}, 
which is slower than the corresponding mean field prediction $\braket{n}_{\mathrm{MF}}(t) \sim (\Gamma t)^{-1}$
(note the different rescaling of time). 

The opposite regime, $\Gamma/\Omega \ll 1$, is the \textit{reaction-limited} one. Here, spatial fluctuations are irrelevant as fast motion makes the particle density homogeneous in space. For classical systems \cite{hinrichsen2000non,vladimir1997nonequilibrium,tauber2005applications,Redner1984,fastdiffusion1992} this regime is described by law of mass action rate equations, which assert that the rate of change of reactants 
is proportional to the product of their global densities. This approach disregards spatial correlations among particles and it indeed reproduces the mean-field result $\braket{n}_{\mathrm{MF}}(t) \sim (\Gamma t)^{-1}$. In what follows, we consider the quantum analogue of this regime, see Fig.~\ref{fig:sketch_RD_GGE}(b)-(c). %, which occurs when the hopping rate is much larger than the reaction rate. 
As we show, this regime is much richer than its classical counterpart, as  coherent effects give rise to collective behavior and quantum correlations beyond mean field.

\textbf{Reaction-limited TGGE}--- For our quantum RD models, the reaction limited regime $\Gamma/\Omega \ll 1$ is equivalent to a weak dissipation limit, which can be analyzed with the recently proposed time-dependent generalized Gibbs ensemble (TGGE) of Refs.~\cite{tGGE1,tGGE2,tGGE3,tGGE4}. Due to fast hopping, one can consider the state of the system $\rho(t)$ to be relaxed with respect to the stationary manifold of the Hamiltonian, $[H,\rho(t)]=0$, at any time $t$. The dynamics of $\rho(t)$ within this manifold is set by the timescale $\Gamma^{-1}$ and it is determined by the dissipation. This aspect is pictorially shown in Fig.~\ref{fig:sketch_RD_GGE}(b). The TGGE approach then makes an ansatz among the set of relaxed states of the Hamiltonian, which is the GGE, see, e.g., Refs.~\cite{GGErev1,GGErev2}. 
In the specific case of the Hamiltonian \eqref{eq:free_fermion_Hamiltonian}, the GGE can be written as 
\begin{equation}
\rho_{\mathrm{GGE}}(t)= \frac{1}{\mathcal{Z}(t)}\mbox{exp}\left(-\sum_{k}\lambda_k(t)\hat{n}_k\right),
\label{eq:tGGE_free_fermions}
\end{equation}
where $\mathcal{Z}(t)=\prod_{k}[1+e^{-\lambda_k(t)}]$. The GGE state \eqref{eq:tGGE_free_fermions} describes averages $\braket{\dots}_{\mathrm{GGE}}(t)$ of local observables in the thermodynamic limit. It is entirely fixed from the knowledge of the Lagrange multipliers $\lambda_{k}(t)$ or, equivalently, of the occupation functions $\braket{\hat{n}_q}_{\mathrm{GGE}}(t)=C_q(t)$, which obey the equations \cite{lossth1,lossth2,lossth3,lossth6}
\begin{equation}
\frac{\mbox{d} C_q(t)}{\mbox{d}t} = \sum_{j,\nu} \braket{{L_j^{\nu}}^{\dagger}[\hat{n}_q,L_j^{\nu}]}_{\mathrm{GGE}}(t),\,\quad  \forall q.
\label{eq:tgge_rate_equation_general}
\end{equation}
The solution $C_q(\tau)$ of this equation clearly depends on the rescaled time $\tau=\Gamma t$, consistently with the above discussion on the reaction-limited regime. The equation of motion \eqref{eq:tgge_rate_equation_general} describes the large-scale dynamics of the system and it has a structure akin to the Boltzmann equation. The right hand side %is named collision term and it 
can be, crucially, exactly computed in the GGE state \eqref{eq:tGGE_free_fermions} through Wick's theorem. To explore the impact of quantum-coherent effects on the RD dynamics, we consider two different initial conditions for Eq.~\eqref{eq:tgge_rate_equation_general}. The first is the coherent Fermi-sea (FS) state with density-filling $0 <n_0\leq 1$: $C_q(t=0)=1$ if $q\in[-\pi n_0,\pi n_0]$, and zero otherwise. The second is the incoherent state $\rho_0 =\mbox{exp}(-\lambda N)/\mathcal{Z}_0$, with a flat initial distribution in momentum space, $C_q(0)=n_0$.   
\begin{figure*}[t]
    \centering
    \includegraphics[width=2\columnwidth]{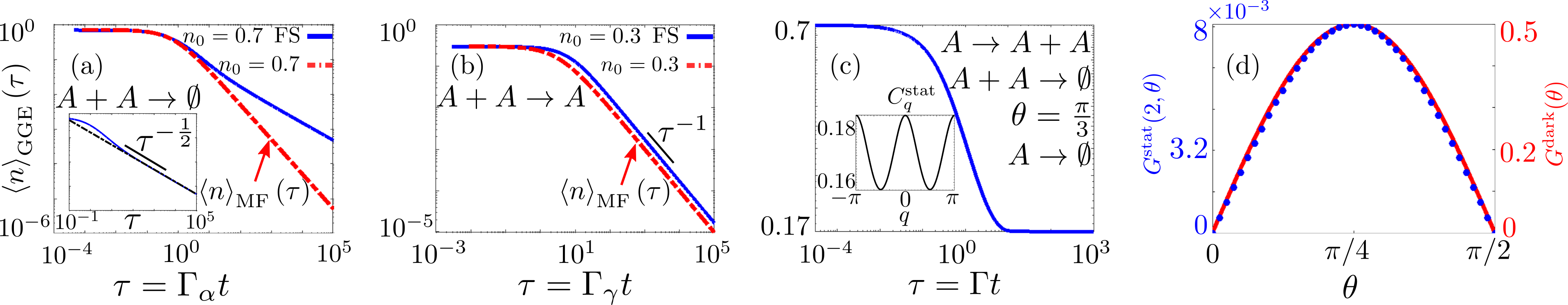}
    \caption{\textbf{Dynamics and active phase in quantum reaction-limited RD systems.} (a) Log-log plot of the particle density $\braket{n}_{\mathrm{GGE}}(\tau)$ as a function of the rescaled time $\tau=\Gamma_{\alpha} t$ for the binary annihilation reaction \eqref{eq:annihilation} with $\theta=0$. In the top-blue curve, the initial state is the coherent Fermi sea (FS) state with filling $n_0=0.7$. The density decays asymptotically as a power law $\braket{n}_{\mathrm{GGE}}(\tau) \sim \tau^{-1/2}$. In the inset, the black dashed curve is a power-law fit $\braket{n}_{\mathrm{GGE}}(\tau)=a\tau^{-b}$ performed over the time window $\tau \in [10^6,10^7]$, with the resulting fitting parameter for the exponent being $b=0.50025\pm 5\cdot 10^{-5}$. In the red-dashed curve, the initial state is the incoherent state $\rho_0$ with the same mean density $n_0=0.7$. In this case, the density is exactly described by the mean-field (MF) law of mass action and $\braket{n}_{\mathrm{GGE}}(\tau)= \braket{n}_{\mathrm{MF}}(\tau) \sim \tau^{-1}$. (b) Log-log plot of the density of particles $\braket{n}_{\mathrm{GGE}}(\tau)$ as a function of $\tau=\Gamma_{\gamma} t$ for the coagulation reaction \eqref{eq:coagulation}. The top-blue curve corresponds to the FS initial state at filling $n_0=0.3$, while the red-dashed one to the incoherent state $\rho_0$ at the same filling. For the FS state, the asymptotic exponent $\braket{n}_{\mathrm{GGE}}(\tau) \sim \tau^{-1}$ is the same as in MF. (c) Log-log plot of the density as a function of $\tau=\Gamma t$ for the CP with pair annihilation Eqs.~\eqref{eq:annihilation}-\eqref{eq:death} and $\Gamma_{\gamma} = 0$, from the FS initial state at $n_0=0.7$. For $\beta>\delta$ an active stationary state is reached. The associated stationary momentum distribution function $C_q^{\mathrm{stat}}$ is shown in the inset as a function of $q$. (d) Stationary correlations $G^{\mathrm{stat}}(2,\theta)$ at distance $2$ (left, blue axis) and dark state contribution $G^{\mathrm{dark}}(\theta)=\sin(2\theta)/2$ (right, red axis) in the CP as a function of $\theta$. Parameters are $\beta=\alpha=1$, $\delta=0.5$.}
     \label{fig:density_plots}
\end{figure*}
 
\textbf{Annihilation and coagulation}---
In Fig.~\ref{fig:density_plots}(a), we plot, from Eq.~\eqref{eq:tgge_rate_equation_general} \cite{SM}, the particle density as a function of time for the pair annihilation reaction only ($\Gamma_{\gamma}=\Gamma_{\beta}=\Gamma_{\delta}=0$), Eq.~\eqref{eq:annihilation} with $\theta=0$, so that interference effects are excluded. The density decays as $\braket{n}_\mathrm{{GGE}}(\tau=\Gamma_{\alpha} t) \sim (\Gamma_{\alpha} t)^{-1/2}$ for the FS initial state for any filling $n_0 \neq 1$. The $1/2$ decay exponent does not necessarily require considering pure states. It also occurs for initial mixed states with an inhomogeneous in $q$ initial occupation function $C_q(0)$ \cite{SM}. In contrast, for the initial state $\rho_0$ and any  $n_0$, the law of mass action is recovered and the density is exactly given by mean field, $\braket{n}_{\mathrm{MF}}(\tau) \sim (\Gamma_{\alpha} t)^{-1}$. This shows the relevance of coherent effects in the critical dynamics of the model, since the algebraic decay of the density in the reaction-limited regime is not described by the mean-field approximation whenever the initial state is quantum coherent. In the latter case, the decay of the particle density is slower than in the classical counterpart of the model, where only incoherent initial states are possible and the long-time behavior of the density is independent on the initial density $n_0$ \cite{Kroon1993,henkel1995equivalences,henkel1997reaction}. 

In Fig.~\ref{fig:density_plots}(b), we plot the particle density as a function of time for the coagulation reaction only  ($\Gamma_{\alpha}=\Gamma_{\beta}=\Gamma_{\delta}=0$), Eq.~\eqref{eq:coagulation} \cite{SM}. %by specializing Eqs.~\eqref{eq:tgge_rate_equation_general} to this case \cite{SM}. 
We find that $\braket{n}_{\mathrm{GGE}}(\tau=\Gamma_{\gamma} t) \sim (\Gamma_{\gamma} t)^{-1}$ both for the incoherent state $\rho_0$ and for the FS state. 
For all initial conditions, we see mean-field like decay
\footnote{For the FS initial state, initial coherences approximately rescale time by an $n_0$ dependent factor without altering the asymptotic mean-field decay.
}, which is different from the situation for pair annihilation at $\theta=0$, Fig.~\ref{fig:density_plots}(a). This difference between annihilation and coagulation processes
is in stark contrast with classical RD models, where 
both processes belong to the same universality class and decay in the same way  independently of initial conditions \cite{fluctuationseffects,Spouge1988,Privman1994,henkel1995equivalences,krebs1995finite,simon1995concentration,henkel1997reaction,ben2005relation}.

For quantum RD, only when starting from the incoherent initial state $\rho_0$ annihilation and coagulation behave in a similar way. In fact, the densities $\braket{n}_{\mathrm{GGE}}^{\mathrm{ann}}(\tau,n_0)$ and $\braket{n}_{\mathrm{GGE}}^{\mathrm{coag}}(\tau,n_0)$ obey
\begin{equation}
\braket{n}_{\mathrm{GGE}}^{\mathrm{coag}}(\tau,n_0)=2 \braket{n}_{\mathrm{GGE}}^{\mathrm{ann}}(\tau,n_0/2),   
\label{eq:coag_ann_relation}
\end{equation}
for $\Gamma_{\alpha}=\Gamma_{\gamma}$. Equation \eqref{eq:coag_ann_relation} is proved noting that the dynamics from the incoherent state $\rho_0$ according to Eq.~\eqref{eq:tgge_rate_equation_general} remains at all times fully incoherent and the quantum master equation \eqref{eq:master_equation} can then be mapped onto a classical master equation \cite{SM}. For the coherent FS initial state, off-diagonal elements of the density matrix $\rho(t)$ are relevant, the quantum master equation does not reduce to its classical counterpart, and Eq.~\eqref{eq:coag_ann_relation} does not apply. This shows that the quantum RD annihilation and coagulation processes do not generically belong to the same universality class and they can display different asymptotic behavior. 

\textbf{Contact process}---
We now consider the contact process (CP) with pair annihilation, cf. Eqs.~\eqref{eq:annihilation}-\eqref{eq:death} with $\Gamma_{\nu}=\Gamma \nu$ ($\nu=\alpha,\beta,\delta$) and $\Gamma_{\gamma} = 0$ and Fig.~\ref{fig:sketch_RD_GGE}(c). 
In Fig.~\ref{fig:density_plots}(c), we plot the density as a function of the rescaled time $\tau=\Gamma t$. We find a phase transition between an absorbing and an active state: the stationary-state density $\braket{n}_{\mathrm{GGE}}^{\mathrm{stat}}$ becomes non-zero when $\beta >\beta_c$, with $\beta_c=\delta$ independent of $\alpha$ and $\theta$. This $\beta_c$ is the same as that of the mean-field classical CP \cite{hinrichsen2000non,henkel2008non}. Furthermore, we find that the associated critical exponents for the stationary density $\braket{n}_{\mathrm{GGE}}^{\mathrm{stat}} \propto (\beta -\beta_c)^{1}$ and for the decay of the density at the critical point $\beta_c$, $\braket{n}_{\mathrm{GGE}}\sim (\Gamma t)^{-1}$, are those of the (mean-field) directed percolation universality. 

Interestingly, however, the stationary state is strongly affected by the quantum coherence introduced by the annihilation reaction in Eq.~\eqref{eq:annihilation}, beyond what can be predicted by a mean-field approach. The inset of Fig.~\ref{fig:density_plots}(c) shows that the different quasi-momenta $q$ are not evenly populated in the stationary state. This applies %both to the initial FS and to the incoherent $\rho_0$ states, 
when $\theta \neq 0,\pi/2$. The non-trivial structure of $C_q^{\mathrm{stat}}$ implies that the stationary state has spatial correlations. To quantify this, we compute the two-point fermionic correlation function $G^{\mathrm{stat}}(x-y,\theta)=\braket{c^{\dagger}_x c_y}_{\mathrm{GGE}}^{\mathrm{stat}}$, which for the mean-field (product) state would be zero unless $x=y$. We find that $G^{\mathrm{stat}}(l,\theta)$ is non zero at even distances $l=2,4,6 \dots$ with a dominant contribution at $l=2$. The value of $G^{\mathrm{stat}}(2,\theta)$ as a function of $\theta$ is shown in Fig.~\ref{fig:density_plots}(d) and is approximately equal to $A(\theta)=\varepsilon\sin(2\theta)/2$. Considering only these dominant next-to-nearest-neighbor correlations, we can identify the (approximate) Lagrange multipliers $\lambda_q^{\rm stat}$ for the stationary GGE $\rho_{\mathrm{GGE}}^{\rm stat}$ expanding to first order in $A(\theta)$ (since $\varepsilon$ is small as shown in Fig.~\ref{fig:density_plots}(d)). One obtains $\lambda^{\rm stat}_q=\lambda_{\rm MF}+\lambda_2\cos (2q)$ and therefore $\rho_{\mathrm{GGE}}^{\rm stat}\propto e^{-\lambda_{\rm MF}N-\lambda_2 Q_2/2}$, with $Q_2=\sum_{j} (c_j^\dagger c_{j+2}+c_{j+2}^\dagger c_j)$.
The contribution $\lambda_{\rm MF}=\log(1/\braket{n}^{\rm stat}_{\mathrm{GGE}}-1)$ represents the mean-field component of the state, while $\lambda_2=-A(\theta)/[\braket{n}^{\rm stat}_{\mathrm{GGE}}(1-\braket{n}^{\rm stat}_{\mathrm{GGE}})]$ accounts for deviations from it. %Expanding, in turn, $\rho_{\mathrm{GGE}}^{\rm stat}$ to first order in $Q_2$  
We show \cite{SM} that $\rho_{\mathrm{GGE}}^{\rm stat}$ can be written in terms of an incoherent state plus a coherent correction, where projectors onto the \textit{local dark states}
\begin{equation}
\ket{\psi}^{\mathrm{dark}, \circ/\bullet}_j=\pm \cos \theta \ket{\bullet (\circ/\bullet)_j \circ}+\sin \theta \ket{\circ (\circ/\bullet)_j \bullet},  
\label{eq:dark_state_annihilation}
\end{equation}
emerge out of the uncorrelated mean-field state. The states $\ket{\psi}^{\mathrm{dark}, \circ/\bullet}_j$ are both dark with respect to the annihilation process \eqref{eq:annihilation} centered in $j$, i.e., $L_j^{\alpha}(\theta)\ket{\psi}^{\mathrm{dark}, \circ/\bullet}_j=0$. Moreover, $\ket{\psi}^{\mathrm{dark}, \bullet}_j$ is dark to branching \eqref{eq:branching} in $j$ and is connected through one-body annihilation \eqref{eq:death} in $j$ to the state $\ket{\psi}^{\mathrm{dark}, \circ}_j$. These local dark states determine the correlations $G^{\mathrm{stat}}(2,\theta)$ in $\rho_{\mathrm{GGE}}^{\rm stat}$, as shown in Fig.~\ref{fig:density_plots}(d).

\textbf{Summary}---
We provided a fully analytical treatment of quantum many-body RD systems in their reaction-limited regime, where the irreversible reaction rates are much smaller than the coherent hopping rate. While for classical RD models this regime is well described by a mean-field approach, we have shown that quantum RD displays instead much richer behaviour. In particular, for annihilation, quantum coherence in the initial state can give rise to an algebraic density decay whose power-law exponent differs from the mean-field one. Furthermore, we have shown that quantum annihilation and coagulation do not belong to the same universality class. For the contact process plus pair annihilation, we have found that the stationary state can feature correlations, which emerge as a consequence of destructive interference. This inherently quantum feature gives rise to locally protected and correlated dark states. The RD systems discussed here connect  the soft-matter physics of chemical reactions to that of cold atoms, where reactions translate into dissipative particle losses or creations  \cite{lossexp1,lossexp2,lossexp3,lossexp4,lossexp5,lossexp6,lossexp7,lossth7,lossth8,lossth1,lossth2,lossth3,lossth4,lossth5,lossth6}, which can be implemented via Rydberg dressing \cite{EXPKConstraint2016,EXPabs2017,EXP2021epidemic}. %Our formulation with the tGGE and the Boltzmann-like equation can be directly generalized to quantum fermionic or bosonic gases in the presence of trapping potential. 
Quantum reaction-diffusion systems are an ideal benchmark to investigate the impact of quantum effects on large-scale universal properties via numerical methods \cite{carollo2019,gillman2019,jo2021} %and quantum simulators \cite{quantum_simulator_1,quantum_simulator_2,quantum_simulator_3,quantum_simulator_4},
and dynamical Keldysh-field theory \cite{kamenev2023field,sieberer2016keldysh}.

\textbf{Acknowledgements.}--- G.P. acknowledges support from the Alexander von Humboldt Foundation through a Humboldt research fellowship for postdoctoral researchers.
We acknowledge financial support in part from EPSRC Grant No.\ EP/R04421X/1, EPSRC Grant No.\ EP/V031201/1, and the Leverhulme Trust Grant No.\ RPG-2018-181. We are also grateful for financing from the Baden-W\"urttemberg Stiftung through Project No.~BWST\_ISF2019-23 and for funding from the Deutsche Forschungsgemeinsschaft (DFG, German Research Foundation) under Project No. 435696605, as well as through the Research Unit FOR 5413/1, Grant No. 465199066. F.C. is indebted to the Baden-W\"urttemberg Stiftung for financial support by the Eliteprogramme for Postdocs.

\bibliography{biblio}

\setcounter{equation}{0}
\setcounter{figure}{0}
\setcounter{table}{0}
\renewcommand{\theequation}{S\arabic{equation}}
\renewcommand{\thefigure}{S\arabic{figure}}

\makeatletter
\renewcommand{\theequation}{S\arabic{figure}}
\renewcommand{\thefigure}{S\arabic{figure}}

\onecolumngrid
\newpage

\setcounter{page}{1}

\setcounter{secnumdepth}{3}
\pagestyle{plain}

\begin{center}
{\Large SUPPLEMENTAL MATERIAL}
\end{center}
\begin{center}
\vspace{0.8cm}
{\Large Reaction-limited quantum reaction-diffusion dynamics in dissipative chains}
\end{center}
\begin{center}
Gabriele Perfetto$^{1}$, Federico Carollo$^{1}$, Juan P. Garrahan$^{2,3}$, and Igor Lesanovsky$^{1,2,3}$
\end{center}
\begin{center}
$^1${\em Institut f\"ur Theoretische Physik, Universit\"at T\"ubingen, Auf der Morgenstelle 14, 72076 T\"ubingen, Germany}\\
$^2${\em School of Physics, Astronomy, University of Nottingham, Nottingham, NG7 2RD, UK.}\\
$^3${\em Centre for the Mathematics, Theoretical Physics of Quantum Non-Equilibrium Systems,
University of Nottingham, Nottingham, NG7 2RD, UK}
\end{center}

\setcounter{equation}{0}
\setcounter{figure}{0}
\setcounter{table}{0}
\setcounter{page}{1}
\makeatletter
\renewcommand{\theequation}{S\arabic{equation}}
\renewcommand{\thefigure}{S\arabic{figure}}

\makeatletter
\renewcommand{\theequation}{S\arabic{equation}}
\renewcommand{\thefigure}{S\arabic{figure}}

\renewcommand{\bibnumfmt}[1]{[S#1]}
\renewcommand{\citenumfont}[1]{S#1}

\onecolumngrid

\setcounter{secnumdepth}{3}

\noindent In this Supplemental Material we provide details about the calculations presented in the main text. In Sec.~\ref{supp:sec_I}, we discuss the TGGE ansatz for the free fermionic hopping Hamilonian, given by Eqs.~(8) and (2) of the main text, respectively. In Sec.~\ref{supp:sec_II}, we specialize the discussion of the reaction-limited TGGE dynamics to the annihilation, coagulation and contact process reactions. In Sec.~\ref{supp:sec_III}, we eventually prove the mapping between the annihilation and the coagulation dynamics in Eq.~(10) of the main text. 

\section{Reaction-limited TGGE ansatz for the fermion hopping Hamiltonian}
\label{supp:sec_I}
We consider the free-fermionic hopping Hamiltonian in (2). We take henceforth periodic boundary conditions $c_{j+L}=c_j$. This choice is without loss of generality as we always consider the thermodynamic limit $L \to \infty$, where the choice of boundary conditions does not matter. The Hamiltonian is diagonalized by Fourier transform \cite{franchini2017introduction}
\begin{equation}
H=-2\Omega \sum_{k_n}\cos(k_n) \hat{n}_{k_n} \qquad \mbox{with} \qquad \hat{n}_k=\hat{c}^{\dagger}_k \hat{c}_k,
\label{supeq:free_fermion_H_diagonal}
\end{equation}
and the operators $\hat{c}_k$ in Fourier space defined as
\begin{equation}
\hat{c}_{k_n}=\frac{1}{\sqrt{L}}\sum_{j=1}^{L} e^{-i k_n j} c_{j}, \qquad \mbox{with inverse} \qquad c_j=\frac{1}{\sqrt{L}}\sum_{k_n}e^{i k_n j} \hat{c}_{k_n}.
\label{supeq:Fourier_transform}
\end{equation}
Here $k_n=2 \pi n/L$, with $n=1,2\dots L$, are the quasi-momenta. In the remainder of this Supplemental Material, we denote summations over the quasi momenta $\sum_{k_n}$ as $\sum_{k}$ for simplicity. The Hamiltonian in Eq.~\eqref{supeq:free_fermion_H_diagonal} clearly commutes with $\hat{n}_k$ for every $k$ value: $[H,\hat{n}_k]=0$. The Hamiltonian is integrable and it possesses an extensive number of conserved charges. The latter are linearly related to the $\hat{n}_k$ operators, see, e.g., the discussion in Refs.~\onlinecite{GGErev1,GGErev2}. The generalized-Gibbs ensemble $\rho_{\mathrm{GGE}}$, describing the relaxation at long times under the unitary dynamics of Eq.~\eqref{supeq:free_fermion_H_diagonal}, can be therefore written in terms of the $\hat{n}_k$ as in Eq.~(8) of the main text. In the reaction limited/weak dissipation regime $\Gamma/\Omega \ll 1$, one promotes the GGE to be time dependent $\rho_{\mathrm{GGE}} \to \rho_{\mathrm{GGE}}(t)$, as proposed in Refs.~\onlinecite{tGGE1,tGGE2,tGGE3,tGGE4}. It is then convenient to introduce the adimensional time $\tau=\Gamma t$, in terms of which the TGGE ansatz is formulated as 
\begin{equation}
\lim_{\Gamma/\Omega \to 0}\rho(t=\tau/\Gamma)=\rho_{\mathrm{GGE}}(\tau)= \frac{1}{\mathcal{Z}(\tau)}\mbox{exp}\left(-\sum_{k}\lambda_k(\tau)\hat{n}_k\right), \qquad \mbox{with} \qquad\frac{\mathrm{d}\rho_{\mathrm{GGE}}(t)}{\mathrm{d}t} = \mathcal{D}[\rho_{\mathrm{GGE}}(t)].
\label{supeq:tGGEevolution}    
\end{equation}
The last equation follows from $[H,\rho_{\mathrm{GGE}}(t)]=0$. We emphasize that the TGGE describes, in the thermodynamic limit $L \to \infty$, the slow evolution taking place on the time scale $\Gamma^{-1}$ of the full quantum state $\rho(t)$. Within this limit, the hopping time $\Omega^{-1}$ (diffusion classically) is much smaller than the reaction time $\Gamma^{-1}$ so that the reactants rapidly mix in space rendering an homogeneous locally in (generalized) equilibrium state $\rho_{\mathrm{GGE}}(\tau)$. The state \eqref{supeq:tGGEevolution} is Gaussian and diagonal in momentum space. Its dynamics is therefore entirely encoded in the two-point function $C_k(\tau)=\braket{\hat{c}_k^{\dagger}\hat{c}_q}_{\mathrm{GGE}}(\tau)=\delta_{k,q}/(\mbox{exp}(\lambda_q)+1) $. In particular, from Eq.~\eqref{supeq:tGGEevolution}, one has 
\begin{equation}
\frac{\mbox{d} C_q(t)}{\mbox{d}t} =\frac{1}{2} \sum_{j,\nu} \braket{{L_j^{\nu}}^{\dagger}[\hat{n}_q,L_j^{\nu}]}_{\mathrm{GGE}}(t)+\braket{[{L_j^{\nu}}^{\dagger},\hat{n}_q]L_j^{\nu}}_{\mathrm{GGE}}(t)= \sum_{j,\nu} \braket{{L_j^{\nu}}^{\dagger}[\hat{n}_q,L_j^{\nu}]}_{\mathrm{GGE}}(t), \quad \forall q.
\label{supeq:tGGEgeneral}
\end{equation}
In the first equality we used the cyclic invariance of the trace, while in the second equality that $[\hat{n}_q,\rho_{\mathrm{GGE}}(t)]=0$. The above equation is recognized as Eq.~(9) of the main text. We notice that the dissipation timescale $\Gamma$ appears as common factor on the right hand side of Eq.~\eqref{supeq:tGGEgeneral} and the solution $C_q(\tau)$ and the TGGE state $\rho_{\mathrm{GGE}}(\tau)$ therefore depend on the rescaled time $\tau$, as anticipated in the main text. 

\section{Quantum reaction-limited dynamics}
\label{supp:sec_II}
In this Section we specialize Eq.~(9) of the main text to the various reaction processes considered. In Subsec.~\ref{supp:sec_II.1}, we consider the binary annihilation reaction (4). In Subsec.~\ref{supp:sec_II.2}, we consider the coagulation reaction (5). In Subsec.~\ref{supp:sec_II.3}, we eventually consider the contact process with binary annihilation in Eqs.~(4), (6) and (7).  

\subsection{Annihilation}
\label{supp:sec_II.1}
For the annihilation reaction $A+A \to \emptyset$ (4), we write the jump operators $L_{j}^{\alpha}$ in Fourier space according to Eq.~\eqref{supeq:Fourier_transform} as
\begin{equation}
L_{j}^{\alpha}(\theta)=\sqrt{\Gamma_{\alpha}}c_j(\cos \theta c_{j+1}-\sin \theta c_{j-1})=\sqrt{\Gamma_{\alpha}}\frac{1}{L}\sum_{k,k'} e^{i(k+k')j}(\cos \theta e^{ik'}-\sin \theta e^{-ik'})\hat{c}_k \hat{c}_{k'}.
\label{supeq:Fourier_annh}
\end{equation}
The following commutation relation is then useful in the evaluation of the commutator in Eq.~(9)
\begin{equation}
[\hat{n}_q, \hat{c}_k \hat{c}_{k'}]=-\hat{c}_k \hat{c}_{k'}(\delta_{k,q}+\delta_{k',q}).
\label{supeq:commutation_annh}
\end{equation}
Inserting Eqs.~\eqref{supeq:Fourier_annh} and \eqref{supeq:commutation_annh} into Eq.~\eqref{supeq:tGGEgeneral} one gets
\begin{equation}
\frac{\mbox{d}C_q(t)}{\mbox{d}t}=-\frac{\Gamma_{\alpha}}{L}\sum_{k,k'} \braket{\hat{c}_{k'}^{\dagger}\hat{c}_{k}^{\dagger}\hat{c}_q \hat{c}_{k+k'-q}}_{\mathrm{GGE}}(t) \left(\cos \theta e^{-ik'}\!-\!\sin \theta e^{ik'} \right)\!\!\left(\cos \theta (e^{i(k+k'-q)}-e^{iq})\!-\sin \theta( e^{-i(k+k'-q)}-e^{-iq})\right).
\label{eq:intermediate_step_annh}
\end{equation}
From the previous equation, it is clear that upon rescaling the time as $\tau=\Gamma_{\alpha} t$, the solution $C_q(\tau)$ depends only on $\tau$. The four-point fermionic function in the previous equation is evaluated exploiting the fact that the TGGE state \eqref{supeq:tGGEgeneral} is Gaussian and therefore Wick theorem applies:
\begin{equation}
\braket{\hat{c}_{k'}^{\dagger}\hat{c}_{k}^{\dagger}\hat{c}_q \hat{c}_{k+k'-q}}_{\mathrm{GGE}}(\tau)=C_k(\tau) C_{k'}(\tau)(\delta_{k,q}-\delta_{k',q}),
\label{supeq:Wick_losses_ab}
\end{equation}
leading to the equation
\begin{equation}
\frac{\mbox{d}C_q(\tau)}{\mbox{d}\tau}=-\frac{1}{L}\sum_k g_{\theta}(k,q) C_k(\tau) C_q(\tau).
\label{eq:ann_rate_equation}
\end{equation}
The function $g_{\theta}(k,q)$ is given by
\begin{align}
g_{\theta}(k,q) &=2(1-\cos(k-q))+\sin (2\theta) (2\cos(k+q)-\cos(2k) -\cos(2q)).
\label{eq:g_function_ann_theta}
\end{align}
Equations \eqref{eq:ann_rate_equation} and \eqref{eq:g_function_ann_theta} have been used with $\theta=0$ to produce the data in Fig.~2(a) of the main text. We checked that the solution of Eqs.~\eqref{eq:ann_rate_equation} and \eqref{eq:g_function_ann_theta} is stable upon increasing $L$ from $L=400, 500$ and $600$ and therefore that the thermodynamic limit is reached. Setting $\theta=0$ into the expression for $g_{\theta}(k,q)$ one has for Eq.~\eqref{eq:ann_rate_equation} that 
\begin{equation}
\frac{\mbox{d}C_q(\tau)}{\mbox{d}\tau}=-2 C_q(t)\braket{n}_{\mathrm{GGE}}(\tau)+\frac{2}{L}\sum_{k}\cos(k-q)C_k(\tau)C_q(\tau),
\label{eq:rate_ann_theta_zero}
\end{equation}
and for the density of reactants \begin{equation}
\braket{n}_{\mathrm{GGE}}(\tau)=\sum_q C_q(\tau)/L,
\label{eq:density_C_q_general}
\end{equation}
that
\begin{equation}
\frac{\mbox{d}\braket{n}_{\mathrm{GGE}}(\tau)}{\mbox{d}\tau}=-2 \braket{n}_{\mathrm{GGE}}^2(\tau)+\frac{2}{L^2}\sum_{k,k'}\cos(k-k')C_k(\tau)C_{k'}(\tau).
\label{eq:rate_ann_theta_zero_density}    
\end{equation}
The last equation is not closed for the density $\braket{n}_{\mathrm{GGE}}(\tau)$ because of the presence of the second term on the right hand side. The first term on the right hand side of \eqref{eq:rate_ann_theta_zero_density} is exactly the mean-field law of mass action describing the reaction-limited regime of classical annihilation RD dynamics \cite{hinrichsen2000non,tauber2005applications,henkel2008non}. The time integration of this contribution simply yields
\begin{equation}
\frac{\mbox{d}\braket{n}_{\mathrm{MF}}(\tau)}{\mbox{d}\tau}=-2\braket{n}_{\mathrm{MF}}^2(\tau) \to \braket{n}_{\mathrm{MF}}(\tau=\Gamma_{\alpha} t)=\frac{n_0}{1+2\Gamma_{\alpha}\,t n_0}.
\label{eq:MF_anndecay}
\end{equation}
The factor $2$ in the previous equation accounts for the fact that in each annihilation reaction $2$ particles are lost. The function $\braket{n}_{\mathrm{MF}}(\tau)$ is depicted in red-dashed in Fig.~2(a). The second term on the right hand side of Eq.~\eqref{eq:rate_ann_theta_zero_density} causes the departure shown in Fig.~2(a) from the law of mass action prediction \eqref{eq:MF_anndecay}. In particular, this term is non-zero if and only if the momentum distribution function $C_q(\tau)$ is not flat in $q$. This is achieved, for example, for the Fermi sea initial state at density $n_0 \neq 1$ and it causes the decay $\braket{n}_{\mathrm{GGE}}(\tau) \sim \tau^{-1/2}$ shown in the blue line of Fig.~2(a). In the opposite case, where $C_q(\tau)$ is flat in the momentum $q$, the second term in the right hand side of \eqref{eq:rate_ann_theta_zero_density} is identically zero and the classical mean-field prediction $\braket{n}_{\mathrm{MF}}(\tau)$ \eqref{eq:MF_anndecay} is exactly retrieved. This is precisely what happens for the incoherent initial state $\rho_0$, with $C_q(\tau=0)=n_0$ for any $q$. 

We mention that the decay of the density in the quantum RD annihilation dynamics ($\theta=0$) has been also studied in Ref.~\onlinecite{RDHorssen} via numerical simulations of quantum-jump trajectories for system sizes up to $L=22$. Therein, the fully occupied initial state is taken, $n_0=1$ with our notation, and the diffusion (hopping)-limited regime $\Omega=\Gamma_{\alpha}$ is considered. The density is found to decay in this limit algebraically as $\braket{n}(t) \sim t^{-b}$, with $1/2<b<1$. In light of our results, we expect the exponent $b(\Omega/\Gamma_{\alpha})$ to vary as a function of $\Omega/\Gamma_{\alpha}$ towards the value $b_{\mathrm{MF}}=1$ in Eq.~\eqref{eq:MF_anndecay} attained in the reaction-limited regime at large $\Omega/\Gamma_{\alpha}$. Similar algebraic decays, with an exponent varying with the Hamiltonian to dissipation strength $\Omega/\Gamma$, have been numerically observed in Refs.~\onlinecite{jo2021,carollo2022quantum} for different types of kinetically-constrained open quantum dynamics.

In the case $\theta \neq 0,\pi/2$, i.e., away from the classical limit of the annihilation reaction, one notices that quantum coherences are produced by the reaction part of the dynamics itself. This translates into the fact that Eq.~\eqref{eq:ann_rate_equation} produces a non-homogeneous momentum occupation function $C_q(\tau)$, even in the case the initial distribution $C_q(0)$ is flat in $q$. As a consequence of this, one observes a decay $\braket{n}_{\mathrm{GGE}}(\tau) \sim \tau^{-1/2}$ for every FS initial state, even at unit filling $n_0=1$, and, more generally, also for the initial state $\rho_0$ at arbitrary $n_0$. This observation is consistent with the results derived in Refs.~\onlinecite{lossth1,lossth2} where two-body atomic losses in one-dimensional bosonic gases in the dissipative quantum Zeno regime have been addressed. 

\subsubsection{Annihilation dynamics from momentum-inhomogeneous GGE initial states}
\begin{figure}[t]
    \centering
    \includegraphics[width=1\columnwidth]{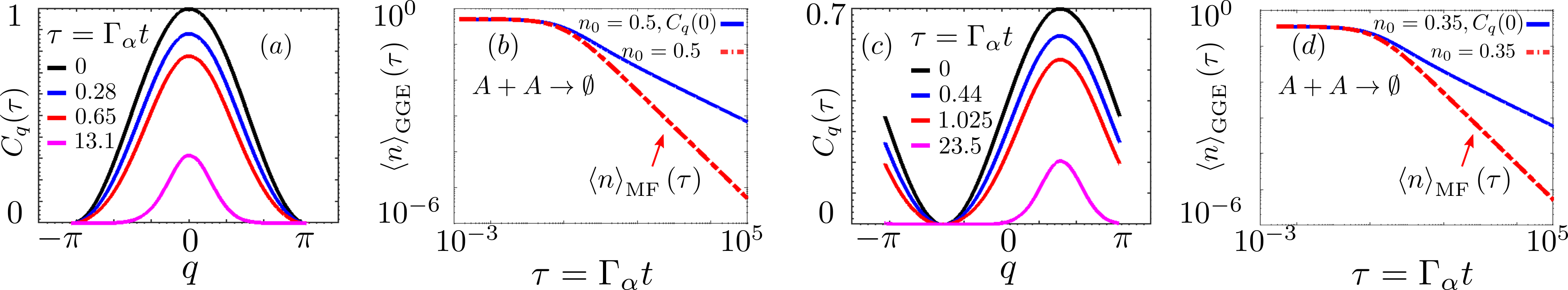}
    \caption{{\textbf{Annihilation dynamics for the occupation function and the particle density.} . (a) Momentum occupation function $C_q(\tau)$ as function of $q$ for increasing values of the rescaled time $\tau=\Gamma_{\alpha}t$ (from top to bottom) for the annihilation reaction at $\theta=0$. The initial occupation function is $C_q(0)=(1+\cos(q))/2$, with initial density $1/2$ (topmost black curve). (b) Top-blue curve: log-log plot of the density $\braket{n}_{\mathrm{GGE}}(\tau)$ associated to the occupation function $C_q(\tau)$ in panel (a). The density decays asymptotically in time as $\braket{n}_{\mathrm{GGE}}(\tau)\sim \tau^{-1/2}$. The bottom red-dashed line gives the mean-field prediction in Eq.~\eqref{eq:MF_anndecay} with $n_0=1/2$ and it is reported for comparison. (c) Momentum occupation function $C_q(\tau)$ as a function of $q$ for increasing values of $\tau$ from the initial condition $C_q(0)=0.7(1+\sin(q))/2$ (topmost black curve), and initial density $n_0=0.35$. Note that in this case $C_q(0)$ is not invariant under quasi-momentum reversal $q\to -q$ and therefore $C_q(\tau)$ evolves differently for positive and negative values of $q$. (d) Top-blue curve: log-log plot of the density $\braket{n}_{\mathrm{GGE}}(\tau)$ associated to $C_q(\tau)$ in (c). The density decays asymptotically as $\braket{n}_{\mathrm{GGE}}(\tau)\sim \tau^{-1/2}$ also in this case, differently from the mean-field prediction (red-dashed curve).}}
    \label{supfig:decay_annihilation_momentum}
\end{figure}

We present here additional analyses and examples in order to further corroborate the results of Subsec.~\ref{supp:sec_II.1} concerning the annihilation decay $\braket{n}_{\mathrm{GGE}}(\tau) \sim \tau^{-1/2}$. Henceforth in this subsection we take $\theta=0$ in Eq.~\eqref{supeq:Fourier_annh}. We consider the case where the initial state has the GGE form in Eq.~\eqref{supeq:tGGEevolution} with momentum inhomogeneous initial Lagrange multipliers $\lambda_{k}(0)$ and momentum occupation function
\begin{equation}
\rho_{\mathrm{GGE}}(0)= \frac{1}{\mathcal{Z}(0)}\mbox{exp}\left(-\sum_{k}\lambda_k(0)\hat{n}_k\right),\quad \mathcal{Z}(0)=\prod_{k_n,n=1}^L \left(1+e^{-\lambda_k(0)}\right), \quad C_q(0)=\frac{1}{\mathrm{exp}(\lambda_q(0))+1}. 
\label{supp:momentum_GGE}
\end{equation}
We call states as in Eq.~\eqref{supp:momentum_GGE} momentum-inhomogeneous GGE initial states as they assume a GGE form and they allow for an initial occupation function $C_q(0)$ not flat in $q$. This implies that the various quasi-momenta $q$ are not uniformly populated in the initial state. It is immediate to compute the purity $\mathcal{P}=\mbox{Tr}[\rho^2]$, for states as in \eqref{supp:momentum_GGE}, as 
\begin{equation}
\mathcal{P}_{\mathrm{GGE}}(0)=\prod_{k_n, n=1}^{L}\mathcal{P}_{k_n}(0), \quad \mbox{with}\quad \mathcal{P}_{k}(0)=\frac{\mbox{Tr}(e^{-2\lambda_k(0)\hat{n}_k})}{\left(1+e^{-\lambda_k(0)}\right)^2}=\frac{1+e^{-2\lambda_k(0)}}{1+2 e^{-\lambda_k(0)} +e^{-2\lambda_k(0)}}.
\label{supp:purity_GGE}
\end{equation}
It is clear that $0\leq\mathcal{P}_k(0)\leq 1$, with the upper bound $1$ attained only if $\lambda_k(0) \to \pm \infty$. From the relation in Eq.~\eqref{supp:momentum_GGE} between $C_q(0)$ and $\lambda_q(0)$, this implies $C_q(0) \to 1$ when $\lambda_q(0)\to -\infty$, and $C_q(0) \to 0$ when $\lambda_q(0) \to +\infty$. Consequently, $\mathcal{P}_{\mathrm{GGE}}(0)<1$, and the initial state is mixed, whenever the occupation function $C_q(0)$ is such that $0<C_q(0)<1$ for at least one quasi-momentum $q$. This is the case, for example, of the state $\rho_0=\mbox{exp}(\lambda N)/\mathcal{Z}_0$, considered in the main text, having flat occupation function: $C_q(0)=n_0<1$ for every $q$. Conversely, for pure states, with $\mathcal{P}_{\mathrm{GGE}}(0)=1$, the occupation function can attain only the values $0$ or $1$. The latter is precisely the limiting case of the Fermi-sea initial pure state. Investigating states of the form \eqref{supp:momentum_GGE} therefore generalizes the analysis of the main text by allowing to study mixed states with a momentum inhomogeneous initial distribution, with the (pure) coherent Fermi sea and the incoherent (momentum-homogeneous) $\rho_0$ states retrieved as particular cases. In Fig.~\ref{supfig:decay_annihilation_momentum}, we show the dynamics of the momentum occupation function $C_q(\tau)$ and the particle density $\braket{n}_{\mathrm{GGE}}(\tau)$ for the annihilation reaction, with $\theta=0$, starting from two different mixed states \eqref{supp:momentum_GGE} with momentum inhomogeneous initial distribution. In both cases, the density decays as $\braket{n}_{\mathrm{GGE}}\sim \tau^{-1/2}$. This shows that the non mean-field decay $\braket{n}_{\mathrm{GGE}}\sim \tau^{-1/2}$ does not necessarily require purity equal one for the initial state (as it is the case for the Fermi sea). Mixed initial states \eqref{supp:momentum_GGE} with purity $\mathcal{P}_{\mathrm{GGE}}(0)<1$ give the decay $\braket{n}_{\mathrm{GGE}}\sim \tau^{-1/2}$ as well, as long as the initial momentum occupation function $C_q(0)$ is inhomogeneous in $q$. The latter is the necessary condition for the decay exponent to be $1/2$, as commented in Subsec.~\ref{supp:sec_II.1} on the basis of Eqs.~\eqref{eq:rate_ann_theta_zero}-\eqref{eq:MF_anndecay}.

\subsection{Coagulation}
\label{supp:sec_II.2}
We consider here the symmetric coagulation reaction (5), where the reactions $A+A\to \emptyset +A$ (right coagulation) and $A+A \to A+\emptyset$ (left coagulation) happen with the same rate $\Gamma_{\gamma}/2$. The generalization of the following analysis to the asymmetric coagulation, where right and left coagulation take place with different rates $\Gamma_{\gamma+}$ and $\Gamma_{\gamma-}$, respectively, is straightforward and it does not alter qualitatively our results.

The calculation for the symmetric coagulation process $L_j^{\gamma\pm}$ in Eq.~(5) is more involved than the one for annihilation as it involves three fermion operators. The expression in Fourier space of the jump operators is
\begin{equation}
L_j^{\gamma\pm}= \sqrt{\Gamma_{\gamma}/2} \, c_j n_{j\pm1}= \sqrt{\Gamma_{\gamma}/2} \, c_j \, c_{j\pm1}^{\dagger}c_{j\pm1}=\sqrt{\frac{\Gamma_{\gamma}}{2}}\frac{1}{L^{3/2}}\sum_{k_1,k_2,k_3} e^{i k_1 j} e^{-ik_2(j\pm 1)}e^{ik_3(j\pm 1)}\hat{c}_{k_1}\hat{c}_{k_2}^{\dagger}\hat{c}_{k_3}.
\label{supeq:symmetric_coagulation_Fourier}
\end{equation}
The commutator in Eq.~\eqref{supeq:tGGEgeneral} can be again simplified using the following commutation relations
\begin{align}
[\hat{n}_q,\hat{c}_{k_1}]=-\delta_{k_1,q} \hat{c}_q \hat{n}_q =-\delta_{k_1,q}\hat{c}_q, \quad \mbox{and} \quad
[\hat{n}_q,\hat{c}_{k_2}^{\dagger}\hat{c}_{k_3}]=\hat{c}_{k_2}^{\dagger}\hat{c}_{k_3}(\delta_{k_2,q}-\delta_{k_3,q}).
\label{supeq:commutator_symm_coagulation}
\end{align}
From Eqs.~\eqref{supeq:symmetric_coagulation_Fourier} and \eqref{supeq:commutator_symm_coagulation} one has
\begin{align}
[\hat{n}_q,L_j^{\gamma\pm}]=\sqrt{\frac{\Gamma_{\gamma}}{2}}\frac{1}{L^{3/2}}&\left[\sum_{k,k'}\hat{c}_k \hat{c}_q^{\dagger}\hat{c}_{k'}e^{i k j}e^{-i q(j\pm1)} e^{i k'(j\pm1)}-\sum_{k,k'}e^{ikj} e^{-ik'(j\pm1)} e^{i q (j\pm1)} \hat{c}_k \hat{c}_{k'}^{\dagger} \hat{c}_q\right. \nonumber \\
&\left.-\sum_{k,k'}e^{iqj} e^{-ik(j\pm1)} e^{ik'(j\pm1)}\hat{c}_q \hat{c}^{\dagger}_k \hat{c}_{k'} \right],
\label{supeq:intermediate_commutator_nq_coagulation}
\end{align}
and therefore
\begin{align}
\sum_j (L_j^{\gamma\pm})^{\dagger}[\hat{n}_q,L_j^{\gamma\pm}] &=\frac{\Gamma_{\gamma}}{2}\frac{1}{L^3}\sum_j\sum_{k_1,k_2,k_3}e^{-ik_1 j}e^{i k_2 (j\pm1)} e^{-i k_3 (j\pm1)}\hat{c}^{\dagger}_{k_3} \hat{c}_{k_2}\hat{c}_{k_1}^{\dagger}\left[ \sum_{k,k'}\hat{c}_k \hat{c}_q^{\dagger}\hat{c}_{k'}e^{i k j}e^{-i q(j\pm1)} e^{i k'(j\pm1)} \right. \nonumber \\
&\left.-\sum_{k,k'}e^{ikj} e^{-ik'(j\pm1)} e^{i q (j\pm1)} \hat{c}_k \hat{c}_{k'}^{\dagger} \hat{c}_q -\sum_{k,k'}e^{iqj} e^{-ik(j\pm1)} e^{ik'(j\pm1)}\hat{c}_q \hat{c}^{\dagger}_k \hat{c}_{k'} \right]\nonumber \\
&=\frac{1}{L^2}\left[\sum_{k_1,k_3,k,k'}e^{\pm i(k_1-k)}(\hat{c}_{k_3}^{\dagger}\hat{c}_{k_1-k+k_3+q-k'}\hat{c}_{k_1}^{\dagger}\hat{c}_k \hat{c}_q^{\dagger} \hat{c}_{k'} -\hat{c}_{k_3}^{\dagger} \hat{c}_{k_1+k_3-k+k'-q}\hat{c}_{k_1}^{\dagger}\hat{c}_k \hat{c}_{k'}^{\dagger}\hat{c}_q) \right] \nonumber \\
-&\frac{1}{L^2}\left[\sum_{k_1,k_3,k,k'}e^{\pm i(k_1-q)}\hat{c}_{k_3}^{\dagger}\hat{c}_{k_1+k_3+k-k'-q}\hat{c}_{k_1}^{\dagger}\hat{c}_q \hat{c}_k^{\dagger} \hat{c}_{k'} \right].
\label{eq:intermediate_six_coagulation}
\end{align}
One realizes from equation \eqref{eq:intermediate_six_coagulation}, that in order to proceed further with the calculation we need to compute six-point fermion correlation functions. This is accomplished, similarly as in the case of the annhilation reaction dynamics in Eq.~\eqref{supeq:Wick_losses_ab}, exploiting the Gaussian structure of the TGGE state \eqref{supeq:tGGEevolution} and therefore the Wick theorem. We report the calculation for the first term in the sum on the third line of Eq.~\eqref{eq:intermediate_six_coagulation}:
\begin{align}
\braket{\hat{c}_{k_3}^{\dagger}\hat{c}_{k_1-k+k_3+q-k'}\hat{c}_{k_1}^{\dagger}\hat{c}_k \hat{c}_q^{\dagger} \hat{c}_{k'}}_{\mathrm{GGE}}(\tau)&= C_{k_1}C_q C_{k_3}\delta_{k_1,k}\delta_{q,k'}+C_{k_3}C_{k_1}(1-C_k)\delta_{k_1,k'}\delta_{k,q}+C_{k_3}C_q(1-C_{k_1})\delta_{k_3,k}\delta_{k',q}  \nonumber \\ -C_{k_3}C_{k_1}&(1-C_q)\delta_{k_3,k}\delta_{k_1,k'}+C_{k_3}(1-C_q)(1-C_{k_1})\delta_{k_3,k'}\delta_{k,q}+C_{k_3}(1-C_q)C_{k_1}\delta_{k_3,k'}\delta_{k_1,k}.
\label{eq:wick_six_points_coagulation_1}
\end{align}
The calculation of the TGGE expectation value of the other two terms on the third line of Eq.~\eqref{eq:intermediate_six_coagulation} works similarly and it is not reported for brevity. In the previous equation, the argument of the momentum occupation function $C_q(\tau)$ is the rescaled time $\tau=\Gamma_{\gamma} t$ and it is not reported again for the sake of brevity. Taking the expectation value of Eq.~\eqref{eq:intermediate_six_coagulation} over the time-dependent GGE state, according to Eq.~\eqref{supeq:tGGEgeneral}, and using the result \eqref{eq:wick_six_points_coagulation_1}, after some algebra one obtains the equation
\begin{equation}
\frac{\mbox{d}C_q(\tau)}{\mbox{d}\tau}= \braket{n}_{\mathrm{GGE}}^2(\tau)-2 C_q(\tau) \braket{n}_{\mathrm{GGE}}(\tau) +\frac{2 C_q(\tau)}{L}\sum_k \cos(q-k)C_k(\tau) -\frac{1}{L^2}\sum_{k,k'}\cos(k-k')C_k(\tau) C_{k'}(\tau),    
\label{eq:rate_equation_coagulation}
\end{equation}
where $\tau=\Gamma_{\gamma} t$ and $\braket{n}_{\mathrm{GGE}}$ as in Eq.~\eqref{eq:density_C_q_general}. The differential equation for the latter quantity can be written by summing Eq.~\eqref{eq:rate_equation_coagulation} over all the modes $q$ 
\begin{equation}
\frac{\mbox{d}\braket{n}_{\mathrm{GGE}}(\tau)}{\mbox{d}\tau}= -\braket{n}_{\mathrm{GGE}}^2(\tau)+\frac{1}{L^2}\left(\sum_{k,k'} \cos(k-k')C_k(\tau) C_{k'}(\tau) \right).
\label{eq:density_eq_coagulation}
\end{equation}
Equation \eqref{eq:rate_equation_coagulation} has been solved with $L=600$ to produce the data in Fig.~2(b) of the main text. Equation \eqref{eq:density_eq_coagulation} has the very same structure as Eq.~\eqref{eq:rate_ann_theta_zero_density} for the annihilation reaction dynamics. The difference between the two reaction processes lies, however, in the evolution equation for the momentum occupation function $C_q(\tau)$, cf. Eq.~\eqref{eq:ann_rate_equation} with Eq.~\eqref{eq:rate_equation_coagulation}. The classical reaction-limited mean-field analysis is encoded into the first term on the right hand side of \eqref{eq:density_eq_coagulation}, which corresponds to the law of mass action for the coagulation dynamics. The time integration of this contribution is
\begin{equation}
\frac{\mbox{d}\braket{n}_{\mathrm{MF}}(\tau)}{\mbox{d}\tau}=-\braket{n}_{\mathrm{MF}}^2(\tau) \to \braket{n}_{\mathrm{MF}}(\tau=\Gamma_{\gamma} t)=\frac{n_0}{1+\Gamma_{\gamma}\,t n_0}.
\label{eq:MF_coagulation}
\end{equation}
We notice that in Eq.~\eqref{eq:MF_coagulation} there is no factor $2$ (as in Eq.~\eqref{eq:MF_anndecay} instead) since each coagulation reaction depletes the number of particles by $1$. The second term in Eq.~\eqref{eq:density_eq_coagulation} is beyond the mean-field classical reaction-limited description and it is not zero if and only if the momentum distribution is not flat in momentum space. This is the case of the FS initial state, whose dynamics is shown in blue in Fig.~2(b). In the opposite case, where $C_q(\tau)$ is flat in momentum space, the mean-field solution \eqref{eq:MF_coagulation} is retrieved. This is the case of the dynamics from the initial state $\rho_0$, which we plot in red-dashed in Fig.~2(b). 

\subsection{Contact process with annihilation}
\label{supp:sec_II.3}
In this Subsection we discuss the contact process with pair annihilation given by Eqs.~(4)-(6) without coagulation $\Gamma_{\gamma}=0$. We consider the case of symmetric branching reactions (7), where $A+\emptyset\to A+A$ (right branching) and $\emptyset+A \to A+A$ (left branching) happen with the same rate $\Gamma_{\beta}/2$. The generalization to asymmetric branching is again straightforward and it does not change qualitatively the results we are going to present. We further rescale all the reactions' rates $\Gamma_{\nu}=\Gamma \nu$, with $\nu=\alpha,\beta$ and $\delta$, such that $\Gamma$ sets the overall dissipation rate, while $\alpha$, $\beta$ and $\delta$ encode the relative strength of the three reactions here considered.

The calculation for the branching jump operator $L_j^{\beta\pm}$ is similar to the one explained in Subsec.~\ref{supp:sec_II.2} for the coagulation dynamics. In particular, one has in Fourier space
\begin{equation}
(L_j^{\beta\pm})^{\dagger}=\sqrt{\frac{\Gamma \beta}{2}}\frac{1}{L^{3/2}}\sum_{k_1,k_2,k_3} e^{i k_1 j} e^{-i k_2 (j\pm1)} e^{i k_3(j\pm1)} \hat{c}_{k_1}\hat{c}^{\dagger}_{k_2} \hat{c}_{k_3}
\label{eq:branching_jump_fourier}.
\end{equation}
From Eqs.~\eqref{eq:branching_jump_fourier} the calculation proceeds in a similar way as the one outlined in Subsec.~\ref{supp:sec_II.2}. We report here just the final result for the sake of brevity
\begin{equation}
\frac{\mbox{d}C_q(\tau)}{\mbox{d}\tau}=\beta\left(2\braket{n}_{\mathrm{GGE}}(\tau) -C_q(\tau)-\braket{n}_{\mathrm{GGE}}^2(\tau)+\frac{1}{L^2}\sum_{k,k'}\cos(k-k')C_k(\tau)C_{k'}(\tau)\right)-\delta C_q -\frac{\alpha}{L}\sum_{k}g_{\theta}(k,q) C_k(\tau)C_q(\tau),
\label{eq:rate_equation_CP_ab}
\end{equation}
with $\tau=\Gamma t$ and $\braket{n}_{\mathrm{GGE}}(\tau)$ in Eq.~\eqref{eq:density_C_q_general}. Equation \eqref{eq:rate_equation_CP_ab} has been solved for $L=600$ to produce the data in Fig.~2(c)-(d). It is also instructive to look at the structure of the equation for the density of particles:
\begin{equation}
\frac{\mbox{d}\braket{n}_{\mathrm{GGE}}(\tau)}{\mbox{d}\tau}=\beta \braket{n}_{\mathrm{GGE}}(\tau)(1-\braket{n}_{\mathrm{GGE}}(\tau))-\delta \braket{n}_{\mathrm{GGE}}(\tau)+\frac{1}{L^2}\sum_{k,k'}\left(\beta\cos(k-k')  -\alpha g_{\theta}(k,k')\right)C_k(\tau)C_{k'}(\tau).
\label{eq:rate_equation_CP_ab_density}
\end{equation}
The part of the right hand side which solely depends on the density $\braket{n}_{\mathrm{GGE}}(\tau)$ gives the result one would get within the mean-field treatment of the classical reaction-limited RD dynamics. The terms coupling different Fourier modes $C_k(\tau)$ and $C_{k'}(\tau)$ go beyond the latter description. From Eq.~\eqref{eq:g_function_ann_theta} and \eqref{eq:rate_equation_CP_ab_density}, the mean-field prediction for the stationary density $n_{\mathrm{MF}}^{\mathrm{stat}}$ is readily obtained
\begin{equation}
\frac{\mbox{d}\braket{n}_{\mathrm{MF}}(\tau)}{\mbox{d}\tau}=-2\alpha \braket{n}_{\mathrm{MF}}^2(\tau)-\delta \braket{n}_{\mathrm{MF}}(\tau)+\beta \braket{n}_{\mathrm{MF}}(\tau)(1-\braket{n}_{\mathrm{MF}}(\tau))=0 \to \braket{n}_{\mathrm{MF}}^{\mathrm{stat}}=\frac{\beta-\delta}{\beta+2\alpha},
\label{eq:MF_CP_ab_stat}
\end{equation}
which is defined ($\braket{n}_{\mathrm{MF}}^{\mathrm{stat}}\geq 0$) only if $\beta>\beta_c=\delta$. In the reaction-limited regime, the critical point of the absorbing-state phase transition therefore coincides with the one of the classical CP (with branching (7) and decay (6) only) above its upper critical dimension \cite{hinrichsen2000non,henkel2008non}. The same stationary state is furthermore obtained from the FS and the incoherent initial state $\rho_0$ (the two initial states having the same density $n_0$). The associated stationary density $\braket{n}_{\mathrm{GGE}}^{\mathrm{stat}}$ is, however, strongly affected by the coherences introduced by the annihilation reaction (4) at $\theta \neq 0 (\pi/2)$. In particular, the stationary occupation function $C_q^{\mathrm{stat}}=\lim_{\tau \to \infty}C_q(\tau)$, shown in the inset of Fig.~2(c), is not flat as a function of $q$, which implies that the quantum reaction-limited steady state displays spatial correlations beyond the classical mean-field description. The stationary density achieved at long times in the active phase, $\beta>\beta_c$, is consequently not given by Eq.~\eqref{eq:MF_CP_ab_stat} for $\theta \neq 0 (\pi/2)$. In the case of Fig.~2(c), for example, we find for $\theta=\pi/3$ (the other parameters are reported in the associated caption) that $\braket{n}_{\mathrm{GGE}}^{\mathrm{stat}} \simeq 0.1706$, while Eq.~\eqref{eq:MF_CP_ab_stat} gives $\braket{n}^{\mathrm{stat}}_{\mathrm{MF}}=1/6\simeq 0.167$. In order to investigate further the structure of the stationary state, we compute the stationary correlation matrix $G^{\mathrm{stat}}(x-y,\theta)=\lim_{\tau \to \infty}G(x-y,\theta,\tau)$, with:
\begin{equation}
G(x-y,\theta,\tau)=\braket{c_x^{\dagger}c_y}_{\mathrm{GGE}}(\tau)= \frac{1}{L}\sum_{k,q}e^{i y q -ix k}\braket{c_k^{\dagger}c_q}_{\mathrm{GGE}}(\tau)= \frac{1}{L}\sum_{q}e^{iq(y-x)}C_q(\tau).
\label{eq:Fouerier_transfrom_structure_factor}
\end{equation}
The latter equation is nothing but the Fourier transform of $C_q(\tau)$. Because of translational invariance, $G(x-y,\theta,\tau)$ is a function of the distance $x-y$ between the two sites only. Moreover, since the initial conditions investigated (the Fermi-sea and the incoherent state $\rho_0$) and Eq.~\eqref{eq:rate_equation_CP_ab} are invariant under quasi-momenta $q \to -q$ reversal, $C_q(\tau)$ is at any time an even function of $q$. As a consequence, the correlation matrix $G(l,\theta,\tau)$ is at any time $\tau$ real and symmetric with respect to the origin: $G^{\ast}(l,\theta,\tau)=G(l,\theta,\tau)=G(-l,\theta,\tau)$.

\begin{figure}[t]
    \centering
    \includegraphics[width=0.5\columnwidth]{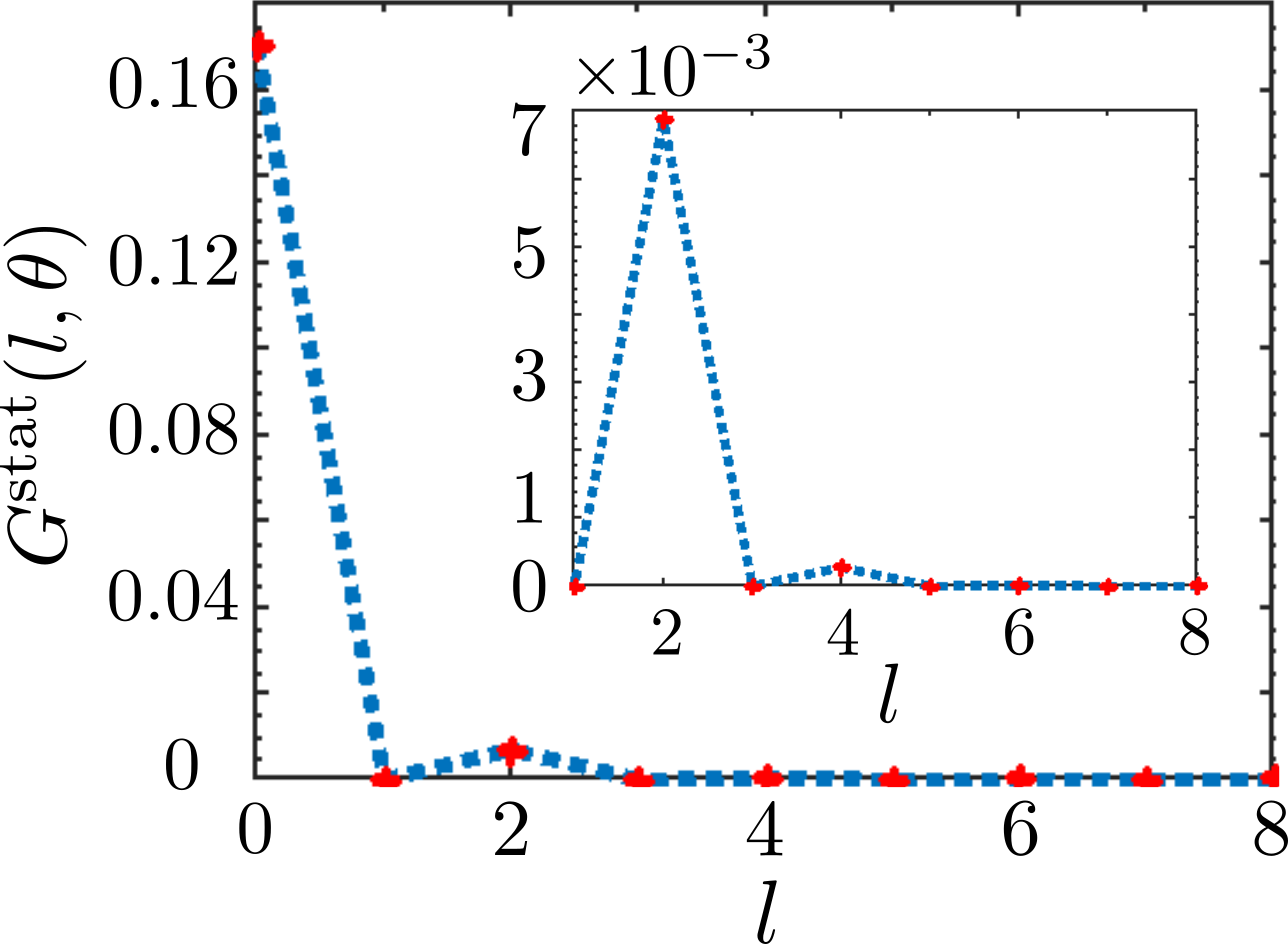}
    \caption{\textbf{Stationary correlations in the CP with binary annihilation.} Plot of $G^{\mathrm{stat}}(l,\theta)$ as a function of $l$ in the active stationary state of the CP with pair annihilation (4) and without coagulation $\Gamma_{\gamma}=0$ (cf. Eqs.~(4)-(6)). The values of $G^{\mathrm{stat}}(l,\theta)$ are represented by the red markers while the blue dashed line is a guide for the eye. The value $G^{\mathrm{stat}}(l=0,\theta)$ at $l=0$ corresponds to the stationary density $\braket{n}_{\mathrm{GGE}}^{\mathrm{stat}}$, which is different from the mean-field value $n_{\mathrm{MF}}^{\mathrm{stat}}$ as long as $\theta \neq 0,\pi/2$. Fundamentally $G^{\mathrm{stat}}(l,\theta)$ is different from zero when $\theta \neq 0,\pi/2$ at even distances $l=\pm 2, 4,6,8,\dots$, showing that the stationary state displays spatial correlations. In the inset, we zoom the values of $G^{\mathrm{stat}}(l,\theta)$ in the interval $l\in[1,8]$, showing that dominant correlations are at distance $l=\pm 2$, while correlations at higher (even) distances are subleading. The parameters are analoguos to the ones used in Fig.~2(c) of the main text: $\beta=\alpha=1$, $\delta=0.5$ and $\theta=\pi/3$. The initial state is the FS at filling $n_0=0.7$. The same result is obtained for the incoherent initial state $\rho_0$ at the same density $n_0$.} 
    \label{supfig:correlation}
\end{figure}

In Fig.~\ref{supfig:correlation}, we plot $G^{\mathrm{stat}}(l,\theta)$ as a function of the distance $l$ between the two sites, with the initial state taken as the FS at filling $n_0=0.7$ (in the same way as in Fig.~2(c)-(d)). One can see that $G^{\mathrm{stat}}(l,\theta)$ has a peak at $l=0$, whose magnitude corresponds to the stationary density $\braket{n}^{\mathrm{stat}}_{\mathrm{GGE}}$ of the active phase. In addition, $G^{\mathrm{stat}}(l,\theta)$ is non-zero at even values of $l=\pm 2, \pm 4, \pm 6\dots$. This fact shows that the stationary active state is not factorized in real space, as it would be within the mean-field description of the classical reaction-limited dynamics. In the inset of Fig.~\ref{supfig:correlation}, we zoom in $G^{\mathrm{stat}}(l,\theta)$ away from $l=0$. The dominant correlations clearly take place at distance $l=2$. Only in the case $\theta=0 \,(\pi/2)$, where the annihilation (4) reduces to its classical limit, the steady state is uncorrelated and one recovers the classical mean-field results: $C_q^{\mathrm{stat}}=\braket{n}^{\mathrm{stat}}_{\mathrm{MF}}$ (flat in momentum space) and $G^{\mathrm{stat}}(l,\theta)$ is zero for any $l\neq 0$.
 
We provide here the explicit expression of the GGE stationary state $\rho_{\mathrm{GGE}}^{\mathrm{stat}}$ in order to better explain the relation between the local dark states in Eq.~(11) of the main text and the non-trivial correlation function displayed in Fig.~\ref{supfig:correlation}. As explained in the main text, the stationary GGE can be written as
$\rho_{\mathrm{GGE}}^{\mathrm{stat}}\propto e^{-\lambda_{\mathrm{MF}}N-\lambda_2 Q_2/2}$, with $\lambda_{\mathrm{MF}}=\log(1/\braket{n}_{\mathrm{GGE}}^{\mathrm{stat}}-1)$ and $\lambda_2=-\varepsilon \sin(\theta)/(2\braket{n}_{\mathrm{GGE}}^{\mathrm{stat}}(1-\braket{n}_{\mathrm{GGE}}^{\mathrm{stat}}))$. We now expand the expression for $\rho_{\mathrm{GGE}}^{\mathrm{stat}}$ to first order in $\varepsilon$ obtaining
\begin{equation}
\rho_{\mathrm{GGE}}^{\mathrm{stat}}=\frac{e^{-\lambda_{\mathrm{MF}}N}}{\mathcal{Z}_{\mathrm{MF}}^{\mathrm{stat}}}-\frac{\lambda_2}{2} \frac{e^{-\lambda_{\mathrm{MF}}N}}{\mathcal{Z}_{\mathrm{MF}}^{\mathrm{stat}}} Q_2 + \mathcal{O}(\varepsilon^2), \,\,  \mbox{with} \,\, Q_2=\sum_{j} (c_j^{\dagger}c_{j+2} +c_{j+2}^{\dagger}c_j),\,\, \mbox{and} \,\, \mathcal{Z}^{\mathrm{stat}}_{\mathrm{MF}}=\mbox{Tr}(e^{-\lambda_{MF}N})=\prod_{j=1}^{L}(1+e^{-\lambda_{\mathrm{MF}}}).      
\end{equation}
In the previous step, we used the fact that $Q_2$ is a conserved charge of the Hamiltonian and therefore $[Q_2,N]=0$ and that $e^{-\lambda_{\mathrm{MF}}N}$ is purely diagonal. The term $e^{-\lambda_{\mathrm{MF}}N}Q_2$ is consequently purely off-diagonal so that the normalization of $\rho_{\mathrm{GGE}}^{\mathrm{stat}}$ is $Z_{\mathrm{MF}}^{\mathrm{stat}}$ to first order in $\varepsilon$. The first term on the right hand side of the equation for $\rho^{\mathrm{stat}}_{\mathrm{GGE}}$ is an incoherent mixture of states in the fermionic Fock space spanned by $\ket{C}=\ket{C_1,C_2\dots C_L}=\ket{\circ_1  \circ_2  \dots \bullet_L}$, with $N\ket{C}=N(C)\ket{C}=\sum_j N(C_j)\ket{C}$, according to the factorized probability measure $\propto e^{-\lambda_{MF} N(C)}$:
\begin{equation}
\frac{e^{-\lambda_{\mathrm{MF}}N}}{\mathcal{Z}_{\mathrm{MF}}^{\mathrm{stat}}}=\sum_{C} \frac{e^{-\lambda_{\mathrm{MF}}N(C)}}{Z_{\mathrm{MF}}^{\mathrm{stat}}}\ket{C}\bra{C}=\sum_{C_1,C_2\dots C_L} \left(\prod_{j=1}^L P_j\right) \ket{C}\bra{C}, \quad \mbox{and} \quad P_j=\frac{e^{-\lambda_{\mathrm{MF}}N(C_j)}}{1+e^{-\lambda_{\mathrm{MF}}}}.
\label{supeq:MF_state}
\end{equation}
One can eventually calculate the action of $Q_2$ onto the state \eqref{supeq:MF_state} which leads to 
\begin{align}
\rho^{\mathrm{stat}}_{\mathrm{GGE}}&=\frac{e^{-\lambda_{\mathrm{MF}}N}}{\mathcal{Z}_{\mathrm{MF}}^{\mathrm{stat}}} -\frac{\braket{n}_{\mathrm{GGE}}^{\mathrm{stat}} A(\theta)}{2}\sum_j\sum_{C_{<j-1},C_{>j+1}} \!\!\!\!\!\!\! M_j(\ket{C_{<j-1}, \circ_{j-1} \bullet_j \bullet_{j+1} ,C_{>j+1}}\bra{C_{<j-1}, \bullet_{j-1}\bullet_j \circ_{j+1}, C_{>j+1}}+\mbox{h.c.}) \nonumber \\
&+\frac{(1-\braket{n}_{\mathrm{GGE}}^{\mathrm{stat}}) A(\theta)}{2}\sum_j\sum_{C_{<j-1},C_{>j+1}} \!\!\!\!\!\!\!M_j(\ket{C_{<j-1}, \circ_{j-1} \circ_j \bullet_{j+1}, C_{>j+1}}\bra{C_{<j-1}, \bullet_{j-1}\circ_j \circ_{j+1} ,C_{>j+1}}+\mbox{h.c.}) \!+\!\mathcal{O}(\varepsilon^2),
\label{supeq:dark_states_GGE_intermediate}
\end{align}
with $A(\theta)=\varepsilon\sin(2\theta)/2$, as defined in the main text. In the previous equation, we denoted with $C_{<j-1}$ ($C_{>j+1}$) the Fock state preceding (following) the site $j-1$ ($j+1$), i.e., $\ket{C_1,C_2\dots C_{j-1}}$ ($\ket{C_{j+2},\dots C_L}$). We have also denoted with $M_j=\prod_{l\neq j,j\pm1}P_l$, the marginal distribution for all the lattice sites but $j-1,j,j+1$. The relation between the previous equation and the local dark states $\ket{\psi}_j^{\mathrm{dark},\circ/\bullet}$ in Eq.~(11) of the main text can be made more explicit upon rewriting Eq.~\eqref{supeq:dark_states_GGE_intermediate} as 
\begin{align}
\rho^{\mathrm{stat}}_{\mathrm{GGE}}=\rho^{\mathrm{stat}}_{\mathrm{diag}}&+\frac{\braket{n}_{\mathrm{GGE}}^{\mathrm{stat}} \varepsilon}{2}\sum_j\sum_{C_{<j-1},C_{>j+1}}\!\!\!\!\!\!\! M_j\ket{C_{<j-1},\psi_j^{\mathrm{dark},\bullet},C_{>j+1}}\bra{C_{<j-1},\psi_j^{\mathrm{dark},\bullet},C_{>j+1}}\nonumber \\
&+\frac{(1-\braket{n}_{\mathrm{GGE}}^{\mathrm{stat}}) \varepsilon}{2}\sum_j\sum_{C_{<j-1},C_{>j+1}}\!\!\!\!\!\!\! M_j\ket{C_{<j-1}, \psi_j^{\mathrm{dark},\circ},C_{>j+1}},\bra{C_{<j-1}, \psi_j^{\mathrm{dark},\circ},C_{>j+1}} +\mathcal{O}(\varepsilon^2)
\label{supeq:dar_states_GGE_1},     
\end{align}
with 
\begin{align}
\rho^{\mathrm{stat}}_{\mathrm{diag}}=& \frac{e^{-\lambda_{\mathrm{MF}}N}}{\mathcal{Z}_{\mathrm{MF}}^{\mathrm{stat}}} -\frac{\braket{n}_{\mathrm{GGE}}^{\mathrm{stat}}\varepsilon \sin^2\theta}{2} \sum_j \sum_{C_{<j-1},C_{>j+1}}\!\!\!\!\!\!\! M_j \ket{C_{<j-1}, \circ_{j-1} \bullet_j \bullet_{j+1} ,C_{>j+1}}\bra{C_{<j-1},\circ_{j-1}\bullet_j \bullet_{j+1} ,C_{>j+1}} \nonumber \\
-&\frac{\braket{n}_{\mathrm{GGE}}^{\mathrm{stat}}\varepsilon \cos^2\theta}{2}\sum_j \sum_{C_{<j-1},C_{>j+1}}\!\!\!\!\!\!\! M_j \ket{C_{<j-1}, \bullet_{j-1} \bullet_j \circ_{j+1},C_{>j+1} }\bra{ C_{<j-1},\bullet_{j-1}\bullet_j \circ_{j+1}, C_{>j+1}} \nonumber \\
-&\frac{(1-\braket{n}_{\mathrm{GGE}}^{\mathrm{stat}})\varepsilon \sin^2\theta}{2} \sum_j \sum_{C_{<j-1},C_{>j+1}}\!\!\!\!\!\!\! M_j \ket{C_{<j-1}, \circ_{j-1} \circ_j \bullet_{j+1},C_{>j+1}}\bra{C_{<j-1}, \circ_{j-1}\circ_j \bullet_{j+1},C_{>j+1}}\nonumber \\
-&\frac{(1-\braket{n}_{\mathrm{GGE}}^{\mathrm{stat}})\varepsilon \cos^2\theta}{2}\sum_j \sum_{C_{<j-1},C_{>j+1}}\!\!\!\!\!\!\! M_j \ket{C_{<j-1}, \bullet_{j-1} \circ_j \circ_{j+1} ,C_{>j+1}}\bra{C_{<j-1}, \bullet_{j-1}\circ_j \circ_{j+1} ,C_{>j+1}}.
\end{align}
The term $\rho_{\mathrm{diag}}^{\mathrm{stat}}$ is incoherent and gives zero contribution to the correlation function, $\mbox{Tr}[c_x^{\dagger}c_y\rho_{\mathrm{diag}}^{\mathrm{stat}}]=0$ for $x\neq y$. The non-trivial correlations in Fig.~2(d) of the main text are entirely determined by the second and third term in Eq.~\eqref{supeq:dar_states_GGE_1} and, in particular, by the coherences introduced by the projectors onto the dark states $\ket{\psi}_j^{\mathrm{dark},\circ/\bullet}$ appearing therein. The non-trivial structure of $\rho_{\mathrm{GGE}}^{\mathrm{stat}}$, determined by the appearance of the conserved charge $Q_2$, is necessarily determined by the dark states $\ket{\psi}_j^{\mathrm{dark},\circ/\bullet}$ of the annihilation reaction. When $\theta=0,\pi/2$ and destructive interference in Eq.~\eqref{supeq:Fourier_annh} is not possible, the dark states are not present and $Q_2$ is as well absent in $\rho_{\mathrm{GGE}}^{\mathrm{stat}}$. The latter is in this case solely determined by the conserved charge $N$ and it is trivially factorized in space and uncorrelated.

\section{Mapping between annihilation and coagulation}
\label{supp:sec_III}
In this Section we discuss for quantum reaction-limited RD systems the mapping between annihilation (4) at $\theta=0$ (or, equivalently, $\pi/2$), and coagulation (5). The mapping is valid for the incoherent initial state $\rho_0$ and it is expressed by Eq.~(10) of the main text, which relates the density of reactants time evolution in the two reaction processes.  

In this Section we use the Jordan-Wigner (JW) transformation to describe the RD dynamics via spin operators  \cite{franchini2017introduction}
\begin{equation}
c_j = S_j \sigma^{-}_j, \quad c_j^{\dagger} = S_j^{\dagger} \sigma^{+}_j, \quad S_j=\prod_{l=1}^{j-1}(-\sigma_l^z), \quad n_j=c_j^{\dagger}c_j=\frac{1+\sigma^z_j}{2}, \quad \sigma^{\pm}_j =\frac{\sigma_j^x \pm i\sigma_j^y}{2},
\label{eq:JW}
\end{equation}
with $\sigma^{x,y,z}_j$ the spin $1/2$ Pauli matrix at site $j$. One realizes that the fermionic number operator 
$n_j=c_j^{\dagger}c_j=\ket{\uparrow}_j\tensor*[_{j}]{\bra{\uparrow}}{}$ is identified with the projector onto the spin up state $\sigma^z_j \ket{\uparrow}_j=+\ket{\uparrow}_j$. The Hermitian operator $S_j=S_j^{\dagger}$ is usually named JW string. The annihilation reaction (4) in terms of the spin operators reads as 
\begin{equation}
L_j^{\alpha}(\theta=0)=
-\sqrt{\Gamma_{\alpha}}\sigma_j^{-}\sigma_{j+1}^{-},
\label{eq:annihilation_spin}
\end{equation}
while the coagulation reaction becomes (5)
\begin{equation}
L_j^{\gamma\pm}=\sqrt{\Gamma_{\gamma}/2} \, S_j \sigma_j^{-}n_{j\pm1}.
\label{eq:coagulation_spin}
\end{equation}
It is important to emphasize that Eq.~\eqref{eq:coagulation_spin} contains the JW string $S_j$ and it is therefore not local in the spin representation. We, however, show in this Section that in the proof of Eq.~(10) the string term $S_j$ in Eq.~\eqref{eq:coagulation_spin} does not matter. The Hamiltonian (2) with the JW transformation becomes the $XX$ spin chain \cite{franchini2017introduction}
\begin{equation}
H=-\Omega \sum_{j=1}^{L} (\sigma_{j}^{-}\sigma_{j+1}^{+}+\sigma_{j}^{+}\sigma_{j+1}^{-}).    
\label{eq:XX_spin_chian}    
\end{equation}
In order to prove Eq.~(10), we introduce also jump operators $L_j^{D,R}$ ($L_j^{D,L}$) giving incoherent hopping to the right (left) 
\begin{equation}
L_j^{D,R}=\sqrt{D} c_{j+1}^{\dagger}c_{j}= \sqrt{D} \sigma_{j+1}^{+}\sigma_{j}^{-}, \quad L_j^{D,L}=\sqrt{D} c_{j}^{\dagger}c_{j+1}= \sqrt{D} \sigma_j^{+}\sigma_{j+1}^{-},
\label{eq:hopping_incoherent}
\end{equation}
at rate $D$. We remark that the boundary terms, $j=L$, in the Hamiltonian \eqref{eq:XX_spin_chian} and the boundary jump operators $L_{j=L}^{\alpha}(\theta=0)$, $L_{j=L}^{\gamma+}$, $L_{j=1}^{\gamma-}$, $L_{j=L}^{D,R}$ and $L_{j=L}^{D,L}$ depend on the parity $(-1)^N$ of the fermionic number $N$. We do not write these terms explicitly here, as the analysis of the reaction-limited regime through the TGGE of Sec.~\ref{supp:sec_I} directly applies in the thermodynamic limit $L\to \infty$. In this limit boundary terms can be neglected. 
In the proof of Eq.~(10), we consider the incoherent initial state $\rho_0$ with mean density $n_0$:
\begin{equation}
\rho_0 =\frac{\mbox{exp}(-\lambda N)}{\mathcal{Z}_0}= \prod_{j=1}^L\left( n_0 n_j   + (1-n_0) (1-n_j)\right) = \prod_{j=1}^L \left(n_0 \ket{\uparrow}_j\tensor*[_{j}]{\bra{\uparrow}}{}+(1-n_0)\ket{\downarrow}_j\tensor*[_{j}]{\bra{\downarrow}}{}\right).  
\label{eq:initial_state_incoherent_spin}
\end{equation}

The reaction-limited dynamics in Eq.~\eqref{supeq:tGGEevolution} from the initial state \eqref{eq:initial_state_incoherent_spin} remains incoherent at all times and diagonal in the classical basis spanned by product states of the form, e.g., $\ket{C}=\ket{\uparrow \uparrow \downarrow \dots \uparrow}$. The reaction-limited Lindblad dynamics \eqref{supeq:tGGEevolution} can be therefore mapped to a classical master equation by introducing the state vector $\ket{P(t)}$:
\begin{subequations}
\begin{align}
\rho(t)&=\sum_{C} P_C(t)\ket{C}\bra{C} \to \ket{P(t)}=\sum_{C}P_C(t)\ket{C}, \\
\frac{\mathrm{d}\rho_{\mathrm{GGE}}(t)}{\mathrm{d}t}&=\mathcal{D}[\rho_{\mathrm{GGE}}(t)] \to \frac{\mathrm{d}P(C,t)}{\mathrm{d}t}=\sum_{C'\neq C}W(C'\to C)P(C',t)-R(C)P(C,t) \to  \frac{\mathrm{d}\ket{P(t)}}{\mathrm{d}t}=-\mathrm{H}\ket{P(t)}.
\end{align}
\label{eq:quantum_classical_mapping}%
\end{subequations}
Here, $\mathrm{H}$ is the Hamiltonian of the classical master equation (not to be confused with the Hamiltonian $H$ ruling the original coherent dynamics (1)-(3)) and it is given by
\begin{equation}
\mathrm{H}=-\sum_{C}\sum_{C'\neq C} W(C'\to C) \ket{C}\bra{C'}+\sum_{C}R(C)\ket{C}\bra{C}.
\label{eq:Hamiltonian_ME_classical}
\end{equation}
Here, the transition $W(C'\to C)$ and the escape rate $R(C)$ are related to the jump operators in the dissipator $\mathcal{D}$ as
\begin{equation}
W(C'\to C) = \sum_j |\braket{C|L_j|C'}|^2, \quad \mbox{and} \quad R(C)=\sum_{C'\neq C}W(C\to C')= \sum_j \braket{C|L_j^{\dagger} L_j|C}.  
\end{equation}
The initial state $\rho_0$ \eqref{eq:initial_state_incoherent_spin} is mapped to the state $\ket{\rho_0}$
\begin{equation}
\rho_0 \to \ket{\rho_0}= \begin{pmatrix} 
n_0 \\
1-n_0
\end{pmatrix}_1 \otimes  \begin{pmatrix} 
n_0 \\
1-n_0
\end{pmatrix}_2 \dots \otimes  \begin{pmatrix} 
n_0 \\
1-n_0
\end{pmatrix}_L.
\label{eq:initial_state_mapping}
\end{equation}
We note that the string operator $S_j$ present in Eq.~\eqref{eq:coagulation_spin} does not contribute to the dynamics for a purely incoherent density matrix, as in Eq.~\eqref{eq:quantum_classical_mapping}, since $S_j^2=1$. For this reason, in the following, we do not consider the JW string $S_j$ in the coagulation jump operators $L_j^{\gamma\pm}$ \eqref{eq:coagulation_spin}.

The mapping to the classical master equation \eqref{eq:quantum_classical_mapping} applies both to the annihilation \eqref{eq:annihilation_spin} $\mathcal{D}_{\alpha}$ and to the coagulation \eqref{eq:coagulation_spin} $\mathcal{D}_{\gamma}$ dissipator. In both the cases, it can be shown that one can include the incoherent hopping \eqref{eq:hopping_incoherent} into the dissipators $\mathcal{D}_{\alpha,D}$ $\mathcal{D}_{\gamma,D}$, as the reaction-limited dynamics in Eqs.~\eqref{eq:rate_ann_theta_zero} and \eqref{eq:rate_equation_coagulation} for $C_q(\tau)$ for the incoherent evolution \eqref{eq:quantum_classical_mapping} is not changed upon including the jump operators \eqref{eq:hopping_incoherent}. The advantage of doing this is that the dissipators $\mathcal{D}_{\alpha,D}$ and $\mathcal{D}_{\gamma,D}$ map under Eq.~\eqref{eq:quantum_classical_mapping} to the classical Hamiltonians $\mathrm{H}_{\alpha,D}^{\mathrm{ann}}$ and $\mathrm{H}_{\gamma,D}^{\mathrm{coag}}$ of the corresponding classical reaction diffusion systems \cite{henkel1995equivalences,krebs1995finite,simon1995concentration,henkel1997reaction,ben2005relation}. In the latter case, the incoherent hopping accounts for the diffusive motion of the classical reactants, with the rate $D$ in Eq.~\eqref{eq:hopping_incoherent} the diffusion constant. For the classical annihilation-diffusion we have
\begin{equation}
\mathrm{H}_{\alpha,D}^{\mathrm{ann}}= -\sum_{j} \left[D( \sigma_{j}^{-}\sigma_{j+1}^{+}+\sigma_{j}^{+}\sigma_{j+1}^{-}) +\frac{\Delta_{\alpha}}{2}\sigma^{z}_j \sigma^z_{j+1}   -\frac{\Gamma_{\alpha}}{4}(\sigma^z_{j}+\sigma^z_{j+1})-\frac{\Delta_{\alpha}-2D}{2}\right]-\Gamma_{\alpha}\sum_j \sigma_j^{-}\sigma_{j+1}^{-},
\end{equation}
with $\Delta_{\alpha}=D-\Gamma_{\alpha}/2$. For the coagulation-diffusion dynamics
\begin{equation}
\mathrm{H}_{\gamma,D}^{\mathrm{coag}}=  -\sum_{j} \left[D( \sigma_{j}^{-}\sigma_{j+1}^{+}+\sigma_{j}^{+}\sigma_{j+1}^{-}) +\frac{\Delta_{\gamma}}{2}\sigma^{z}_j \sigma^z_{j+1}   -\frac{\Gamma_{\gamma}}{4}(\sigma^z_{j}+\sigma^z_{j+1})-\frac{\Delta_{\gamma}-2D}{2}\right]-\Gamma_{\gamma}\sum_j (\sigma_j^{-}n_{j+1}+\sigma^{-}_{j+1}n_j),
\end{equation}
and $\Delta_{\gamma}=D-\Gamma_{\gamma}/2$. At $\Gamma_{\gamma}=\Gamma_{\alpha}$, the two Hamiltonians are related through a similiarity transformation $B$, as shown in Ref.~\onlinecite{krebs1995finite}:
\begin{equation}
\mathrm{H}_{\gamma=\alpha,D}^{\mathrm{coag}}=B \, \mathrm{H}_{\alpha,D}^{\mathrm{ann}} \, B^{-1}, \quad \mbox{with} \quad B= \bigotimes_{j=1}^{L} B_j \quad \mbox{and} \quad B_j=\begin{pmatrix} 2 & 0 \\
-1  & 1
\end{pmatrix}_j, \quad B_j^{-1}=\begin{pmatrix} 1/2 & 0 \\
1/2  & 1
\end{pmatrix}_j.
\label{eq:classical_similarity}
\end{equation}
The similarity matrix $B$ is built as the $L-$fold tensor product of the matrix $B_j$ at the site $j$, which is the same for every lattice site. This equation in the classical realm is considered as the hallmark of the equivalence between annihilation and coagulation \cite{hinrichsen2000non,henkel1995equivalences,henkel1997reaction,krebs1995finite}. Once Eq.~\eqref{eq:classical_similarity} is established, the equivalence between the quantum reaction-limited annihilation dynamics and the coagulation one is readily is established. Namely, one has 
\begin{align}
\braket{n}_{\mathrm{GGE}}^{\mathrm{coag}}(\tau,n_0)&=\mbox{Tr}[n_j \rho_{\mathrm{GGE}}^{\gamma}(t)]=\mbox{Tr}[n_j \rho_{\mathrm{GGE}}^{\gamma,D}(t)]=\braket{-|n_j \, \mbox{exp}(-\mathrm{H}_{\gamma=\alpha,D}^{\mathrm{coag}}t)|\rho_0}= \braket{-|n_j \, B \, \mbox{exp}(-\mathrm{H}_{\alpha,D}^{\mathrm{ann}}t) \, B^{-1}|\rho_0} \nonumber \\
&=2\braket{-|n_j\, \mbox{exp}(-\mathrm{H}_{\alpha}^{\mathrm{ann}}t)|\rho_0/2}=2\braket{n}_{\mathrm{GGE}}^{\mathrm{ann}}(\tau,n_0/2),
\label{eq:mapping_proved}
\end{align}
which corresponds to Eq.~(10) of the main text and it implies that the density of reactants decays with the same exponent in the two processes. Here, $\ket{-}=\sum_C \ket{C}$ is the flat state, which implements the normalization of the classical dynamics. In the second equality, we used the fact that for an incoherent dynamics the introduction of the incoherent hopping does not alter the quantum reaction-limited dynamics. In the third equality, we used the mapping \eqref{eq:quantum_classical_mapping} and, in the fourth equality, Eq.~\eqref{eq:classical_similarity}. In the fifth equality, we used that $n_j B_j=2 n_j$, $\bra{-}B_j=\bra{-}$ and $B^{-1}\ket{\rho_0}=\ket{\rho_0/2}$. The derivation of Eq.~\eqref{eq:mapping_proved} can be straightforwardly extended to density equal-time correlation functions \cite{henkel1995equivalences,krebs1995finite,henkel1997reaction,ben2005relation}.

Equations \eqref{eq:MF_anndecay} and \eqref{eq:MF_coagulation} do satisfy Eq.~\eqref{eq:mapping_proved}. This derivation therefore shows that the quantum reaction-limited dynamics of the annihilation \eqref{eq:annihilation_spin} and coagulation \eqref{eq:coagulation_spin} processes from the initial state \eqref{eq:initial_state_incoherent_spin} is exactly coincident with the mean-field classical reaction-limited evolution. Departures from the latter equivalence are of intrinsic quantum nature and they are determined either by coherences in the initial state, as in the case of the FS initial state, or by coherences introduced by the reactions, as in the case of Eq.~(4) at $\theta \neq 0,\pi/2$. In both these cases, the density matrix $\rho(t)$ develops in time coherences $\propto \ket{C}\bra{C'}$ and the quantum master equation (1) cannot be mapped to its classical counterpart, as in Eq.~\eqref{eq:quantum_classical_mapping}. In light of our results in Fig.~2(a)-(b), where the density of particles decays with different asymptotic exponents in the two processes, we conclude that Eq.~\eqref{eq:mapping_proved} does not hold when coherences are present and quantum annihilation and coagulation generically display different asymptotic decays (with different power-law exponents). This is in sharp contrast with the classical case where annihilation and coagulation display analogous algebraic decays irrespectively of the initial condition. 

It is important to emphasize that the equivalence between annihilation and coagulation is here proved in the reaction-limited regime in a weak way, i.e., in terms of the relation \eqref{eq:mapping_proved} between the densities in the dynamics of the two processes. In the classical case \cite{hinrichsen2000non,henkel1995equivalences,henkel1997reaction,krebs1995finite}, the annihilation-coagulation equivalence is proved in a stronger way in terms of the similarity relation \eqref{eq:classical_similarity} between the associated dynamical generators $H^{\mathrm{ann}}_{\alpha,D}$ and $H^{\mathrm{coag}}_{\gamma=\alpha,D}$. Our results, however, do not rule out the possibility of the equivalence between quantum annihilation and coagulation, in the stronger sense that the associated Lindbladians (1) (including therefore both the Hamiltonian $H$ in Eq.~(2) and the dissipator $\mathcal{D}$ (3)) are related through a similarity transformation. As a matter of fact, in Ref.~\onlinecite{RDHorssen}, it has been suggested that a similarity transformation relating the Lindbladians of the two process does exist. This conclusion is drawn in Ref.~\onlinecite{RDHorssen} on the basis of exact-numerical diagonalization of the Lindbladian generators of the two processes for up to $L=8$ spins. The existence of such similarity relation between the two Lindblad generators does not, however, generically imply a simple relation, as Eq.~\eqref{eq:mapping_proved}, for the densities $\braket{n}^{\mathrm{ann}}(t)$ and $\braket{n}^{\mathrm{coag}}(t)$ in the two processes and therefore the same algebraic decay. One must, indeed, also consider how the observable $n_j$ and the initial state $B^{-1}\ket{\rho_0}$ transform under the similarity transformation $B$. In order to establish Eq.~\eqref{eq:mapping_proved} it is, indeed, crucial that $B^{-1}\ket{\rho_0}=\ket{\rho_0/2}$, i.e., the initial state is simply transformed to a state of the same form but with halved density. Our results seem to indicate that for the FS initial state the transformation rule is more intricate, though a more in-depth analysis, which we leave for future studies, is needed.

\end{document}